\documentclass[preprint,showpacs,preprintnumbers,amsmath,amssymb,floatfix]{revtex4-1}
\usepackage{graphicx}
\usepackage{dcolumn}
\usepackage{amsmath} \usepackage{bm}
\begin{document}
\preprint{}

\title{Bucket Shaking Stops Bunch Oscillations In The Tevatron }
\thanks{Operated by Fermi Research Alliance, LLC under Contract No. DE-AC02-07CH11359 with the United States Department of Energy.}
\author{C.Y. Tan}
\author{A. Burov}
\affiliation{Fermi National Accelerator Laboratory, P.O. Box 500,
  Batavia, IL 60510-5011, USA.}

\date{\today}

\begin{abstract}
  Bunches in the Tevatron are known to exhibit longitudinal
  oscillations which persist indefinitely. These oscillations are
  colloquially called ``dancing bunches''. Although the dancing
  bunches do not cause single bunch emittance growth or beam loss at
  injection, it leads to bunch lengthening at collisions. In
  operations, a longitudinal damper has been built which stops this
  dance and damps out coupled bunch modes. Recent theoretical work
  predicts that the dance can also be stopped by an appropriate change
  in the bunch distribution. This paper shows the Tevatron experiments
  which support this theory.
\end{abstract}
\maketitle

\section{Introduction}

Ever since the start of Run~II, the bunches in the Tevatron have been
observed to have longitudinal oscillations which persist
indefinitely. The reason for the persistence of these oscillations
have been traced to the loss of Landau damping (LLD) caused by the
inductive impedance of the Tevatron \cite{ronmoore}; these
oscillations are colloquially called "dancing bunches". At the
injection energy of 150~GeV, these oscillations do not seem to yield
any emittance growth or any beam loss. But at the flattop energy of
980~GeV, they lead to an effective bunch length growth which reduces
luminosity. In order to stop the dance, a longitudinal damper system
has been built which damps it out \cite{dampers}.

Recently, theoretical work has predicted that the dance can also be
stopped by flattening out its phase space distribution at low
synchrotron frequencies \cite{burov1, burov2}. In particular, this
flattening can be achieved by shaking the RF phase at the synchrotron
frequency of the low amplitude particles \cite{burov3}. The goal of
this paper is to demonstrate experimentally that the dance can be
stopped by changing the beam distribution appropriately.

\section{Theory}

The Boltzmann-Jeans-Vlasov (BJV) equation \cite{henon} is
conventionally used to describe longitudinal motion of bunched
beams. This equation has a continuous spectrum and, possibly, a
discrete one \cite{kampen, ecker}. The discrete van Kampen modes are
described with regular functions and some of them do not decay.
Therefore, in principle, any coupled bunch wake drives an instability
when there is LLD. However, in practice, the coupled bunch wake has to
be high enough to give an observable growth rate. If the growth rate
is too small, LLD results in persistent oscillations caused by initial
perturbations.

For bunched beams, LLD was first discussed and estimated by F.~Sacherer \cite{sacherer}.
Later, his main results were re-derived and discussed in more details
by other authors \cite{besnier, hofmann, ng, metral, gonzalez}. For a
dipole mode, all the approaches were actually based on an assumption
that the bunch moves as a rigid body.  However, recent solutions of
the eigenvalue problem \cite{burov1, burov2} show that the rigid bunch
approximation can lead to significant over estimations of the LLD threshold.

As it is known since the original Sacherer's paper \cite{sacherer},
the threshold bunch population $N_{\rm th}$ is a strong function of
the bunch length $\ell$. In particular, for an inductive wake above
transition,
\begin{equation}
N_{\rm th}\propto \ell^5 .
\label{threshold}
\end{equation}
This scaling law can be derived from an idea that Landau damping is
lost whenever the incoherent tune shift $\Delta\Omega\propto N
Z_{||}(\ell^{-1})/\ell^2$ exceeds the incoherent tune spread
$\delta\Omega\propto \ell^2$.  For the inductive impedance, the
incoherent tune shift decreases with the bunch length as $\Delta
\Omega \propto \ell^{-3}$. The combined action of this decrease with
increasing nonlinear tune spread $\delta \Omega \propto \ell^2$
results in $\ell^5$ in Eq.~\ref{threshold}.  This high sensitivity on
the bunch length indicates that approximations of the bunch profile or
arbitrary assumptions about the eigenfunctions can lead to significant
errors in the calculated LLD threshold because they can change the
effective bunch length. For example, for a full bucket of a
single-harmonic RF system with an inductive impedance above
transition, the threshold relative tune shift $\Delta \Omega / \Omega$
was found to be as low as 10\% for the Hofmann-Pedersen distribution,
and just $\sim\!  1$\% for a model of the Tevatron coalesced
\footnote{One coalesced bunch means that 7 bunches from the Main
  Injector are ``coalesced'' into one high current bunch and injected
  into the Tevatron. For our experiment, two coalesced bunches
  separated by 21 buckets are used.}  bunch \cite{burov1}.  In terms
of bunch population, the two thresholds differ by almost two orders of
magnitude. It turns out that the onset of LLD is highly sensitive to
the steepness of the distribution function at low amplitudes: the
flatter the distribution, the more stable it is.  This prediction
appears to be generally correct when the bare RF synchrotron frequency
monotonically decreases with amplitude and the wake field is
repulsive, i.e.~the wake lowers the incoherent synchrotron
frequencies.  This conclusion agrees with Ref.~\cite{metral}, where
the LLD threshold was calculated for several distributions with the
inductive impedance above and below transition. It was shown there
that below transition LLD is sensitive to the edges of the
distribution, while above transition, it is sensitive to the flatness
of the bunch core.

As was discussed in Ref.~\cite{burov1}, in the case of a sinusoidal RF
system, any combination of inductance, wall resistivity, or high order
cavity modes above transition, or space charge below transition will
shift the incoherent spectrum down to lower frequency and the coherent
mode emerges above it. Since the incoherent frequencies of low
amplitude particles are close to the mode frequency, their wieght in
the mode dominates. Hence, for a single-harmonic RF and a repulsive
wake function, the discrete mode causes dipole motion of the bunch
center while its tails remain still.  This is the behavior of the
bunches in the Tevatron \cite{ronmoore}.

\subsection{Flattening out the distribution for particles with small
  amplitudes}
\label{flattening.sec}

To flatten out the bunch distribution at small amplitudes, resonant
shaking of the RF phase was suggested for the Tevatron \cite{burov3},
with an idea to use anomalous diffusion within a controlled phase
space area; see Ref.\cite{jeon, huang, syleebook} and references
therein. 

Let it be assumed that the RF phase is modulated at a frequency
$\Omega_m$, which is close to the synchrotron frequency
$\Omega_s$. Let the amplitude of the modulation $\phi_m(t)$
adiabatically grow from zero, then stay a while at some value $\phi_0$
and then adiabatically decrease to zero. To prevent excitation of the
tail particles and the coherent modes, the process has to be
adiabatic. However, even when the process is generally adiabatic,
i.e.~when $|d\phi_m/dt| \ll \Omega_s \phi_0$, the adiabaticity for
some particles is going to be broken. Indeed, resonant RF shaking
results in either one or two stable fixed points (SFPs) inside the
bucket. In the last case, there is an inner separatrix between the two
SFPs and when the modulation amplitude changes, the separatrix moves and
some particles cross it. Separatrix crossing is a non-adiabatic
process resulting in classical chaos and anomalous diffusion.

Thus, the phase space density can be changed only in the case of two
SFPs which occur when the modulation frequency is
lower than the synchrotron frequency, $\Omega_m < \Omega_s$, and the
modulation amplitude is lower than its bifurcation value, $\phi_m <
\phi_b = 3.08\epsilon^{3/2}$ with $\epsilon
=1-\Omega_m/\Omega_s$. When the modulation amplitude grows from zero
to its bifurcation value, and when it comes back to zero later, the
irreversible change of the phase space density occurs for the phase
space area
with action $J \leq J_{\lim}$, where
\begin{equation}
J_{\lim} \approx 6\epsilon J_{\rm{bucket}},
\label{Jlim}
\end{equation}
and $J_{\rm{bucket}}$ is the bucket acceptance. For dimensionless
variables associated with the unperturbed Hamiltonian
$H(z,p)=p^2/2+1-\cos z$, the acceptance $J_{\rm{bucket}}=8/\pi$. The
numerical factor ``6'' in Eq.~(\ref{Jlim}) was approximated from a
numerical solution discussed below and it
is about two times larger than the separatrix border at zero
amplitude. After this adiabatic cycle, the phase space density becomes
nearly constant for the entire area $J<J_{\lim}$, provided that the
shaking amplitude crosses its bifurcation value, i.e.

\begin{equation}
\phi_0 \geq 3.08\epsilon^{3/2}
\label{phi_0}
\end{equation}

It is worth mentioning that the adiabatically ramped modulation does
not excite any coherent motion when the modulation is turned
off. Thus, to make a flat phase space density within a certain action
$J_{\lim}$, the adiabatic RF phase modulation has to be applied
slightly below the synchrotron frequency, $\epsilon =
0.16J_{\lim}/J_{\rm{bucket}}$, and its amplitude must cross the
bifurcation, Eq.~(\ref{phi_0}).

%\subsubsection{Computer Simulations}
%\label{simulations.sec}

A simulation of how the bunch distribution is modified with RF phase
shaking has been done using the following map
\begin{align}\nonumber
z_{n+1} &= z_n + p_n\Delta t \\
p_{n+1} &= p_n - \Delta t\sin\bigg[z_{n+1} - \phi_m(t_n)\sin(1-\epsilon)t_n\bigg]\\
\nonumber
t_{n+1} &= t_n + \Delta t
\label{bs.1}
\end{align}
where $z_n$ and $p_n$ are the coordinate and momenta respectively in
dimensionless units, $t_n$ is the time variable in radians of the
synchrotron oscillation, $\Delta t$ is its numerical step. The
amplitude of the RF phase modulation $\phi_m(t)$ was taken as a
trapezoid similar to Fig.~\ref{ramp.eps}.

Here are the typical parameters used in the simulations:
\begin{itemize}
\item the adiabaticity parameter $\dot{\phi_m}/(\Omega_s \phi_0)\sim
  200$. 
\item $\Delta t=0.01$ radians which is small enough for the results to be
  independent of its specific value. 
\item Initial phase space density is assumed to be $F(J) \propto
  (J_{\rm  max}-J)^2$ with the emittance $J_{\rm max}$ set close to the bucket
  acceptance. 
\item Number of macro-particles $N=4 \times 10^4$.
\end{itemize}

\begin{figure*}[htb]
  \centering
 \includegraphics*[width=120mm,clip]{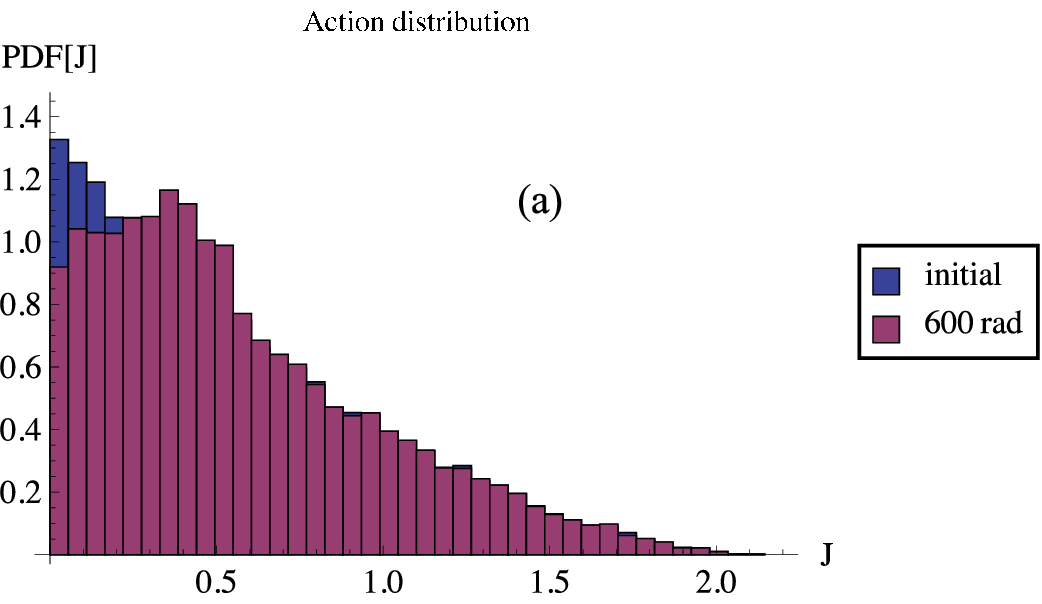}
 \includegraphics*[width=120mm,clip]{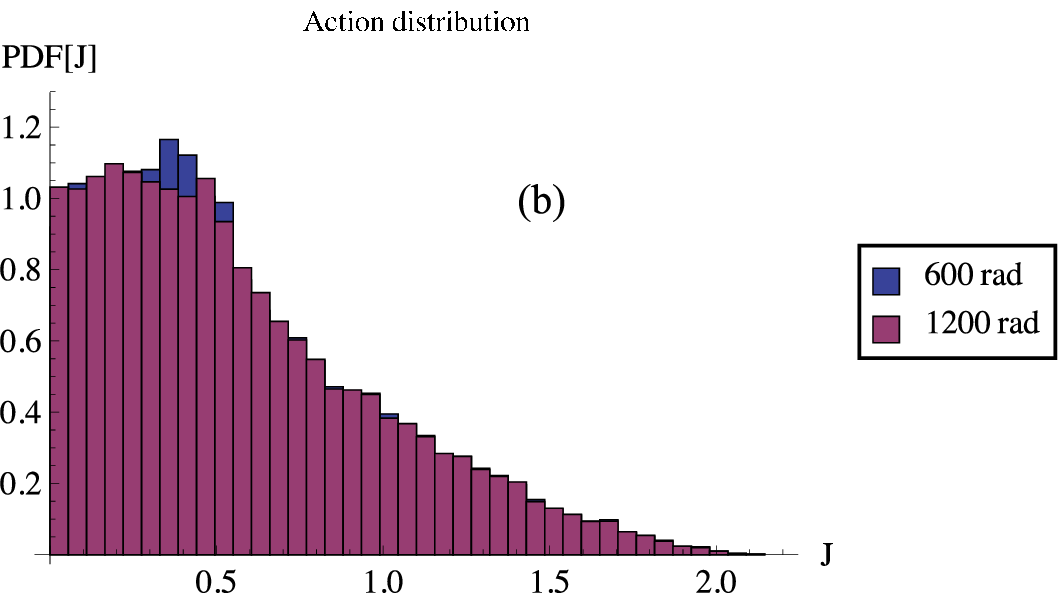}
 \includegraphics*[width=120mm,clip]{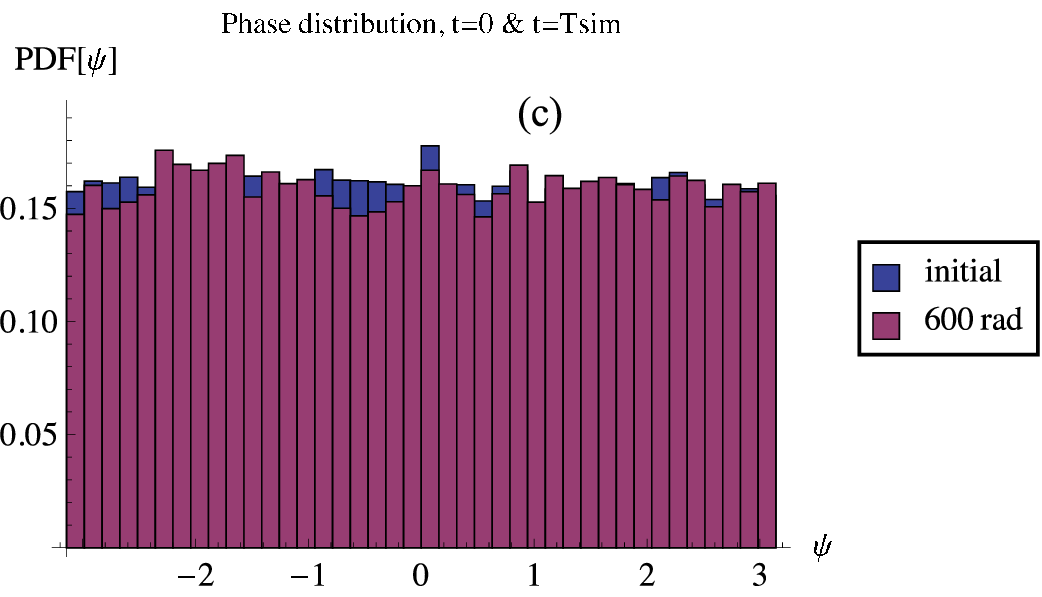}

 \caption{(a) distributions over action {\tt PDF[J]}, original (blue) and after
   the first ramp (pink); the overlapped area is in
   violet. (b) a similar comparison of the distributions before and
   after the second ramp. (c) a comparison of the phase
   distributions {\tt PDF[$\psi$]} before and after the first ramp.}
  \label{PDF12.eps}
\end{figure*}

The simulation results before and after shaking are shown in
Fig.~\ref{PDF12.eps} for $\epsilon=0.03$, $\phi_0=0.025$ and two
consecutive shaking cycles with $T_{\rm{sim}} = 600$ radians or about
90 synchrotron periods each. Each cycle time was equally divided into
three parts of about 30~synchrotron periods each: a linear growth of
the modulation amplitude from 0 to $\phi_0$, staying at $\phi_0$, and
a linear decrease from $\phi_0$ back to 0.

Clearly the action distribution {\tt PDF[J]\/} has
successfully flattened out and there is even a little divot that is less
pronounced after the second shaking cycle. Except for this small
difference, the second cycle does not significantly change the
distribution. The phase distribution {\tt PDF[$\psi$]\/} after every
cycle is as flat as before, showing that no coherent oscillations were
excited. 

\begin{figure*}[htb]
  \centering
 \includegraphics*[width=120mm,clip]{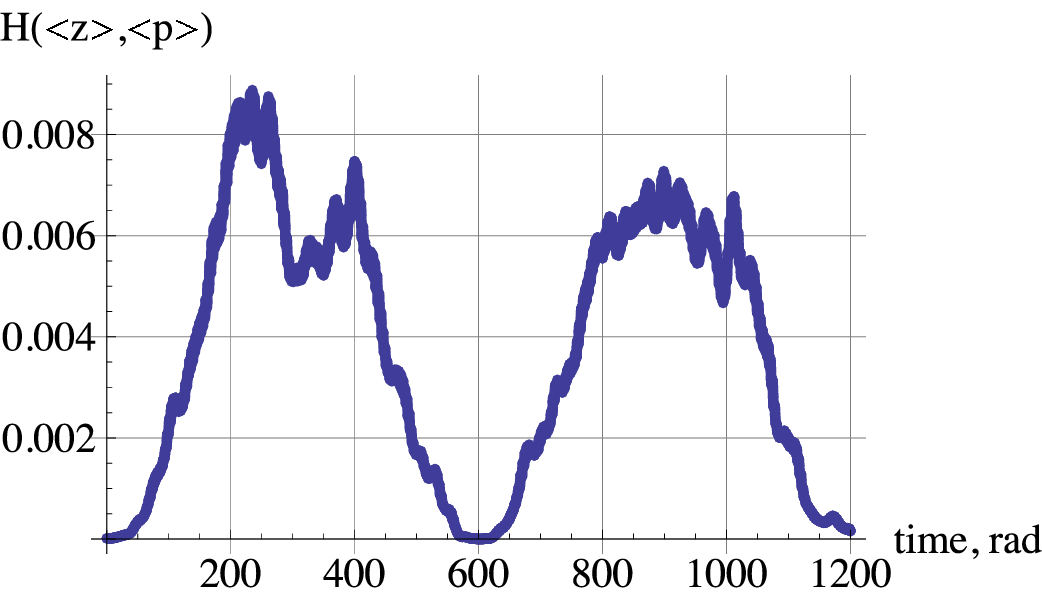}

 \caption{Time dependence of the unperturbed Hamiltonian taken for the bunch-average coordinate and momentum.}
  \label{HamT.eps}
\end{figure*}

The time dependence of the unperturbed Hamiltonian $H(z,p)=p^2/2 +
(1-\cos z)$ calculated for the bunch average values of the canonical
variables $\langle z\rangle$ and $\langle p\rangle$ is shown in
Fig.~\ref{HamT.eps}. This simulation shows that the adiabaticity of
shaking is really important: after every cycle, the Hamiltonian goes
to zero. The irregular features of this plot probably reflect the
chaotic nature of the anomalous diffusion responsible for the
flattening of the distribution.

%\begin{figure*}[htb]
%  \centering
% \includegraphics*[width=120mm,clip]{f1a.eps}
% \includegraphics*[width=120mm,clip]{f1b.eps}
% \caption{Initial (blue) and final (pink) distributions over action
%   and phase. The overlapping area is in violet.}
%  \label{f1f2.eps}
%\end{figure*}

\section{Experiment}

\begin{figure*}[htb]
  \centering
 \includegraphics*[width=120mm,clip]{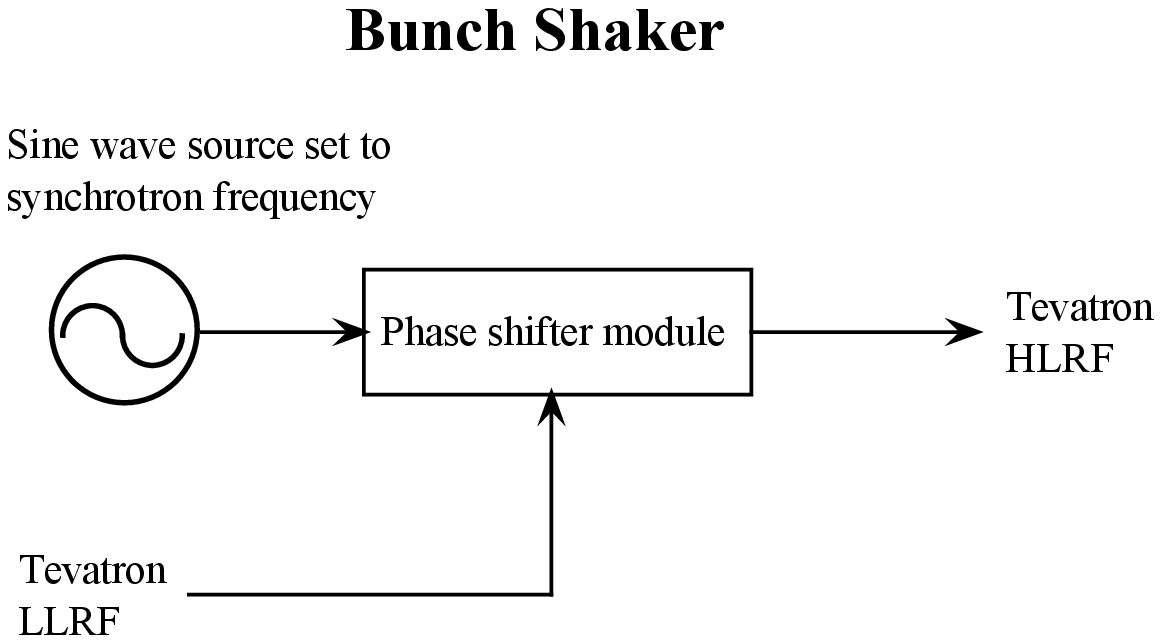}
 \includegraphics*[width=120mm,clip]{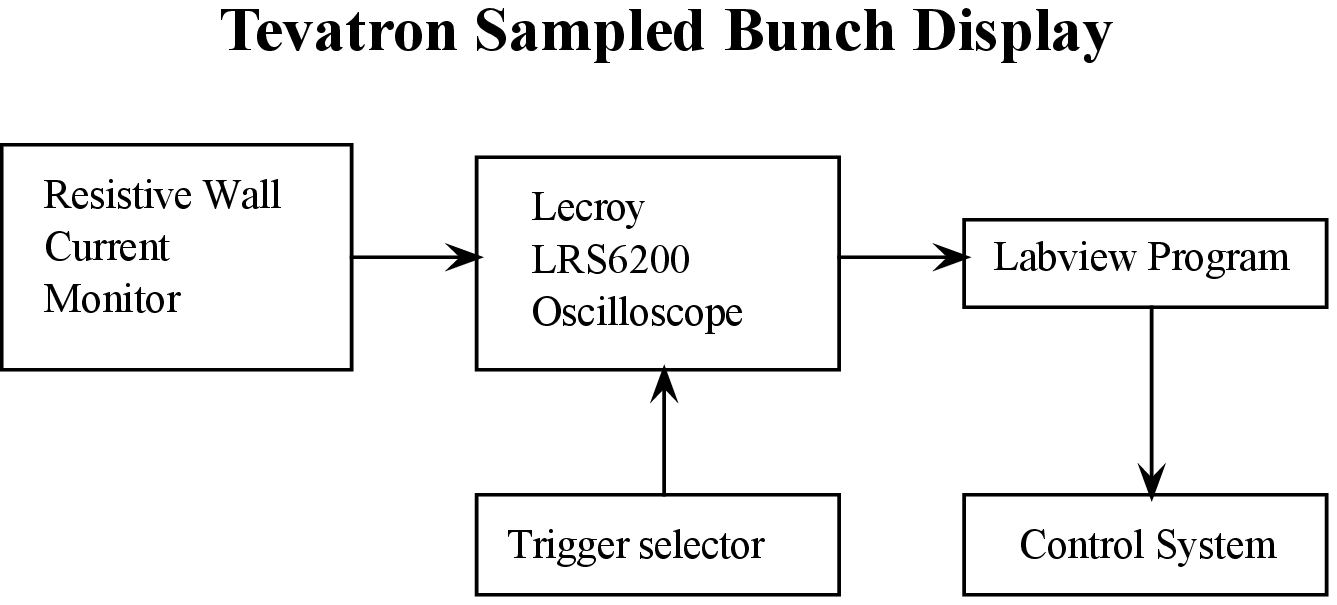}
 \includegraphics*[width=120mm,clip]{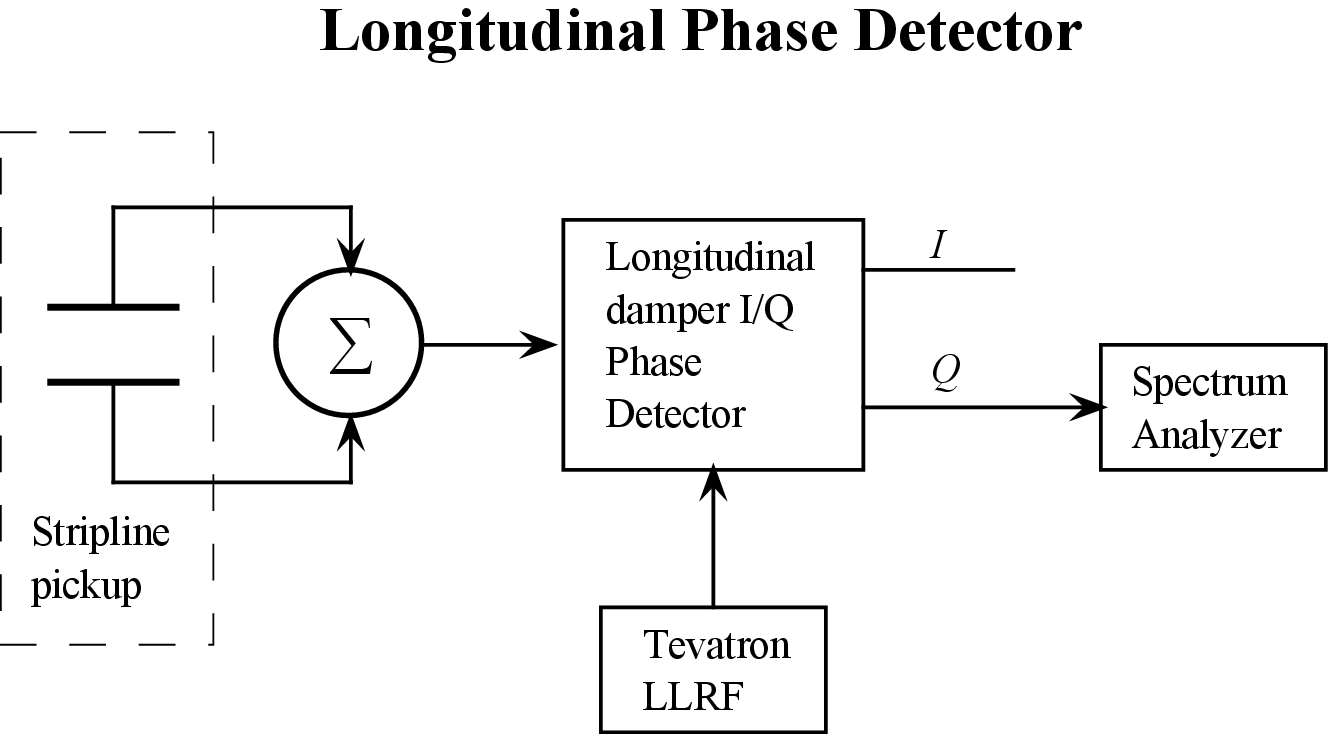}
 \caption{The block diagrams of the bunch shaker and detectors used to
   monitor the bunches for the experiments.}
  \label{setup.eps}
\end{figure*}

The block diagram of the phase modulation hardware used for shaking
the beam is shown in Fig.~\ref{setup.eps}. A signal generator
generates a sine wave where its amplitude and frequency can be
programmed and its output is fed into a phase shifter module. The
phase shifter modulates the Tevatron low level RF (LLRF)
and the result is fed into the Tevatron high level RF (HLRF).
Essentially, the block diagram produces the following
\begin{equation}
f_{\rm HLRF} = A \sin\bigg[2\pi f_{\rm LLRF}t + \phi_m(t)\sin( 2\pi f_mt)
+ \theta\bigg]
\label{expt.1}
\end{equation}
where $f_{\rm HLRF}$ is the phase modulated signal sent to the HLRF, $A$ is
the amplitude of the signal sent to the HLRF, $f_{\rm LLRF}$ is the frequency from the
LLRF, $\theta$ is an arbitrary phase,  and the signal generator produces
the amplitude $\phi_m$ and frequency $f_m$ for the phase modulation.

The time evolution of the bunch during the experiment are measured
using the Sampled Bunch Display (SBD) \cite{pordes}. Its block diagram
is shown in Fig.~\ref{setup.eps}.  The SBD measures the bunch profile
from a resistive wall current monitor with an oscilloscope that has a
2~GHz bandwidth and the collected data is processed with a
LabView program which calculates the following parameters:
\begin{itemize}
\item bunch centroid,
\item  bunch current,
\item rms bunch length.
\end{itemize}
These parameters are then returned to the
control system and can be plotted to give
Figures~\ref{lumber150.eps} and \ref{datalogger.eps}. Furthermore, the
snapshots of the bunch from the resistive wall signal can also be
downloaded. The SBD trigger has been set up to take five consecutive
snapshots of the bunch that are 1~s apart. These snapshots are
presented in the figures below.

The block diagram of the phase detector used to measure the
longitudinal motion of the bunch w.r.t.~the Tevatron RF is shown in
Fig.~\ref{setup.eps}. The $I/Q$ phase detector is a part of the Tevatron
longitudinal damper system \cite{dampers} which essentially takes the sum
signal from a stripline pickup, down-converts it with the Tevatron LLRF and low pass
filters it to produce a quadrature signal. The quadrature signal is then
measured with a spectrum analyzer.

The Tevatron parameters relevant to the experiment are shown in
Table~\ref{tab:table1}. This experiment only uses two coalesced proton
bunches and measurements are either taken at the injection energy of
150~GeV or at the flattop energy of 980~GeV.

\begin{table}[b]
\caption{\label{tab:table1}%
Tevatron Parameters Relevant To The Experiment
}
\centering
\begin{ruledtabular}
\begin{tabular}{ccc}
\textrm{Parameter}&
\textrm{Value}&
\textrm{Units}\\
\colrule
Injection energy & 150 & GeV \\
Flattop energy & 980 & GeV\\
Synchrotron frequency at 150 GeV & $87.47$ & Hz\\
Synchrotron frequency at 980 GeV & $34.75$ & Hz\\
RF frequency at 150 GeV & $53.103$ & MHz\\
RF frequency at 980 GeV & $53.104$ & MHz\\
Harmonic number & 1113 & --\\
Buckets between two injected bunches & 21 & -- \\
Intensity per bunch & $(200 - 300) \times 10^9$ &--\\
\end{tabular}
\end{ruledtabular}
\end{table}

\subsection{Results at the injection energy of 150 GeV}

\begin{figure*}[htb]
  \centering
 {\includegraphics*[width=100mm,clip]{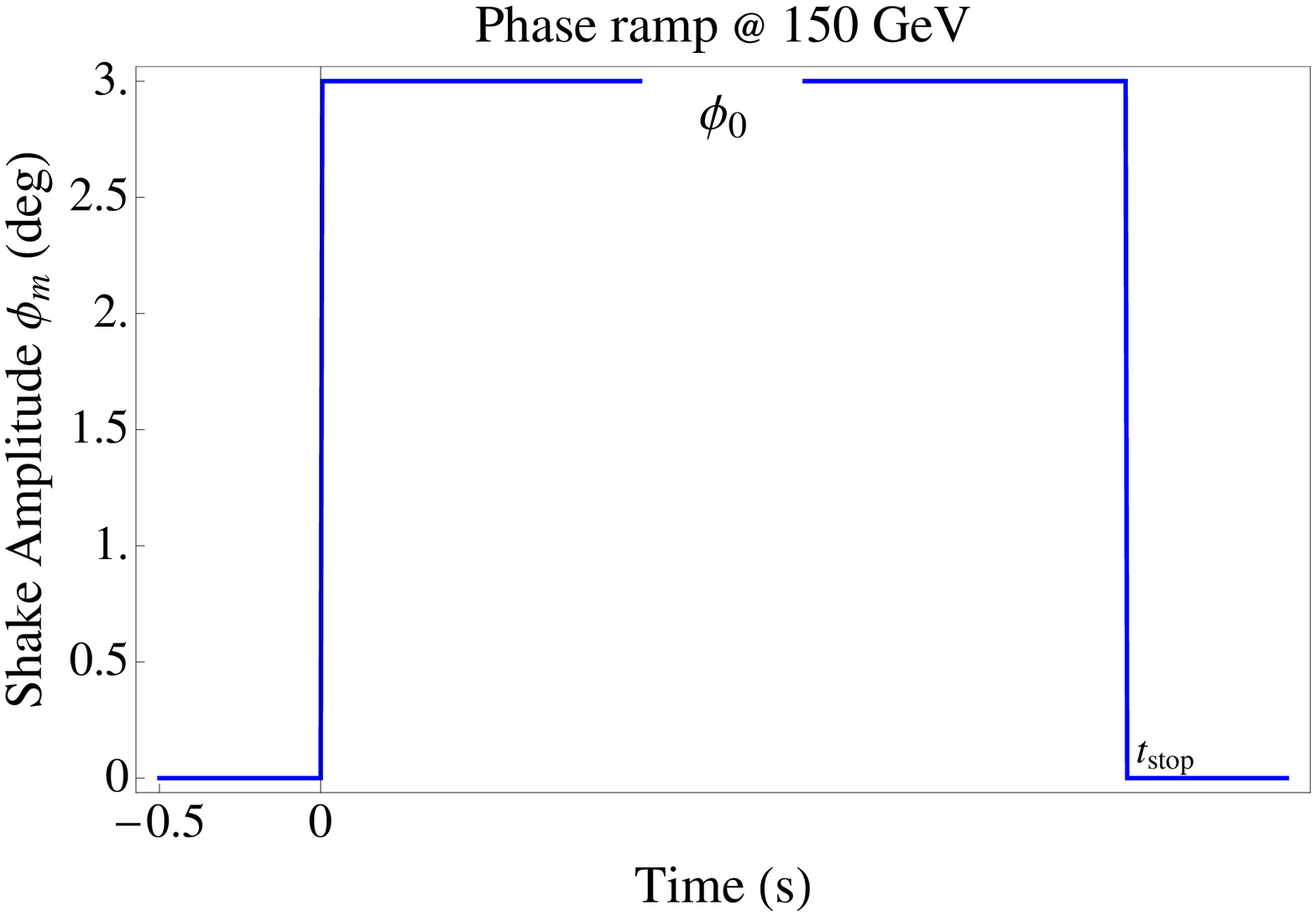}}
 {\includegraphics*[width=100mm,clip]{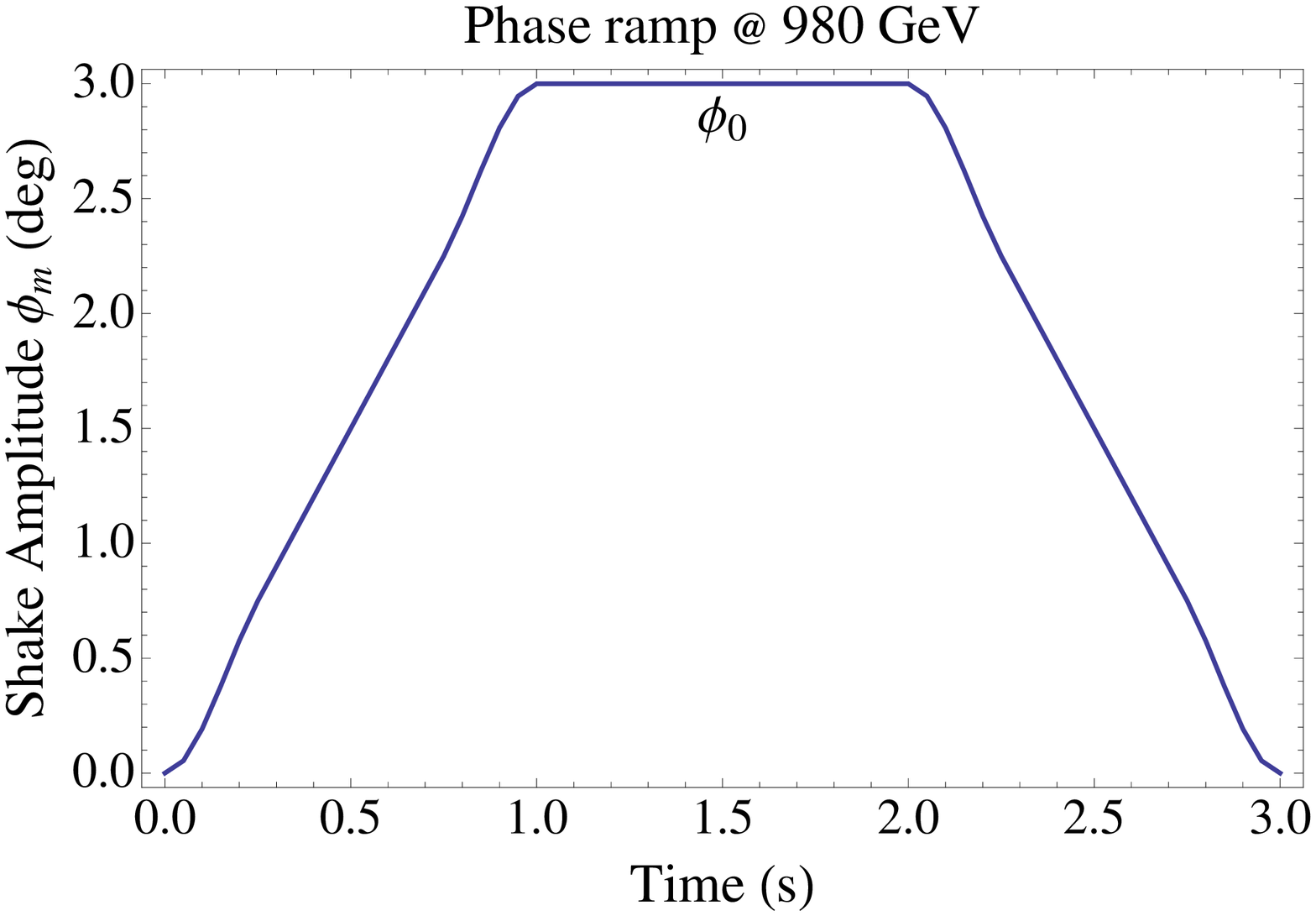}}
 \caption{These are the phase ramps $\phi_m(t)$ used at
   150~GeV and 980~GeV. For the 150~GeV experiments, $t_{\rm stop}$ is user defined.}
  \label{ramp.eps}
\end{figure*}

The results in this section have been performed at the injection energy of
150~GeV. At injection, the bunch nearly fills the bucket and so there
are small beam current losses whenever the bunch is shaken. (Results
at flattop do not have this problem. See section~\ref{980GeV.sec}).
In this experiment, $f_m$ has been set to $87.47$~Hz because it is the
measured synchrotron frequency $f_s$ and the bunch is shaken one time for
14~s. (Note: theoretically, $f_m$ should have been set to a frequency
which is {\it smaller\/} than $f_s$. However, at the time, this
criterion was not appreciated and the experiment was not done.)
The phase ramp used in the 150~GeV experiments is not adiabatic
and is shown in Fig.~\ref{ramp.eps}.  The maximum amplitude of the shake
has been tested for $\phi_0 = 1^\circ$, $2^\circ$ and $3^\circ$
respectively and experimentally, $\phi_0=3^\circ$ has been found to
produce the best effect for the duration of the shake.

\begin{figure*}[htb]
  \centering
 \includegraphics*[width=120mm,clip]{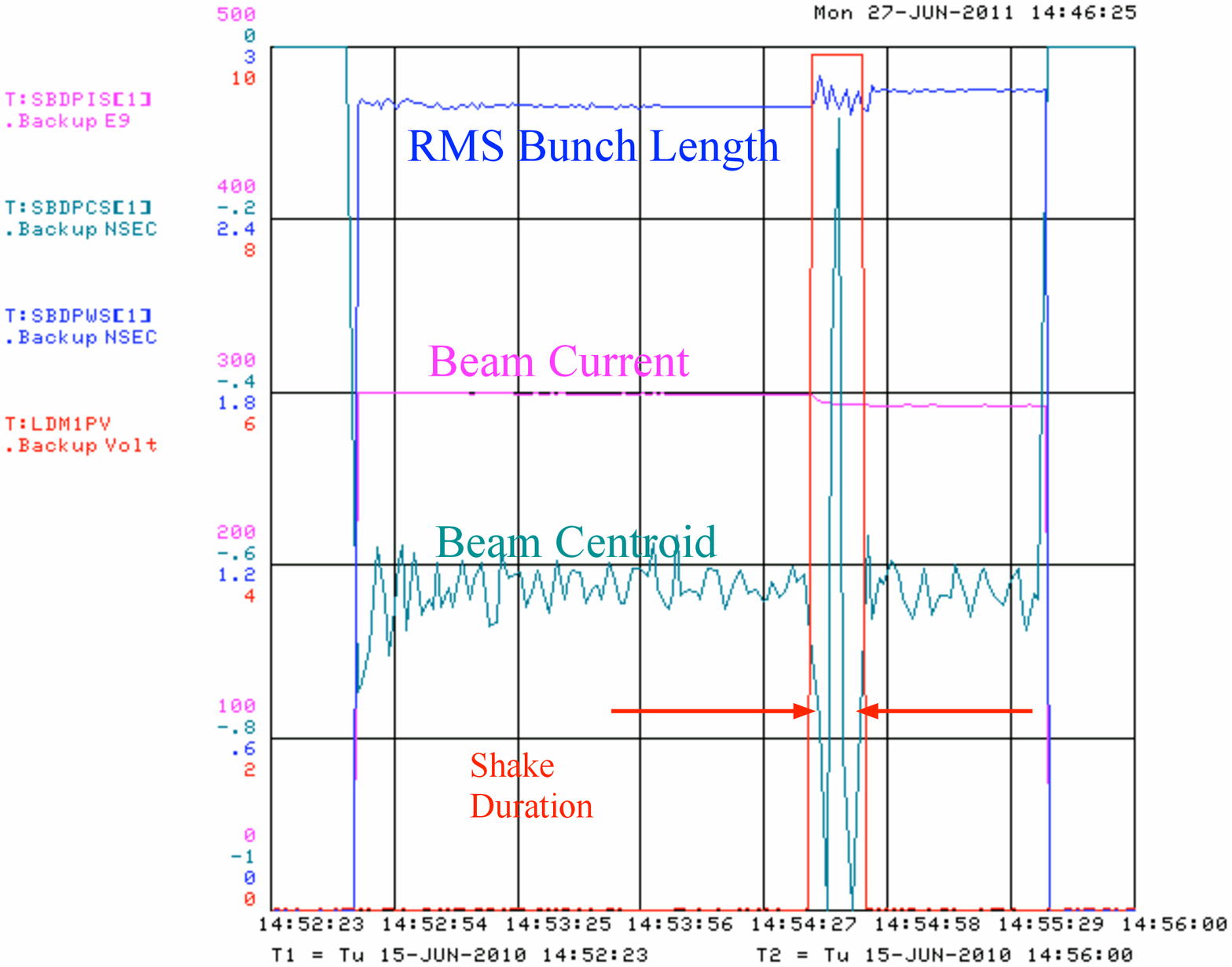}
 \caption{The data from the SBD system is plotted here: T:SBDPWS,
   T:SBDPIS and T:SBDPCS. The beam is shaken for 14~s and there is some beam loss and
   bunch length growth. Although the measured bunch centroid looks like it is
 still oscillating, the snapshots show that the dancing has
 stopped. See Fig.~\ref{beforeafter150.eps}.}
  \label{lumber150.eps}
\end{figure*}

\begin{figure*}[htb]
  \centering
 \includegraphics*[width=120mm,clip]{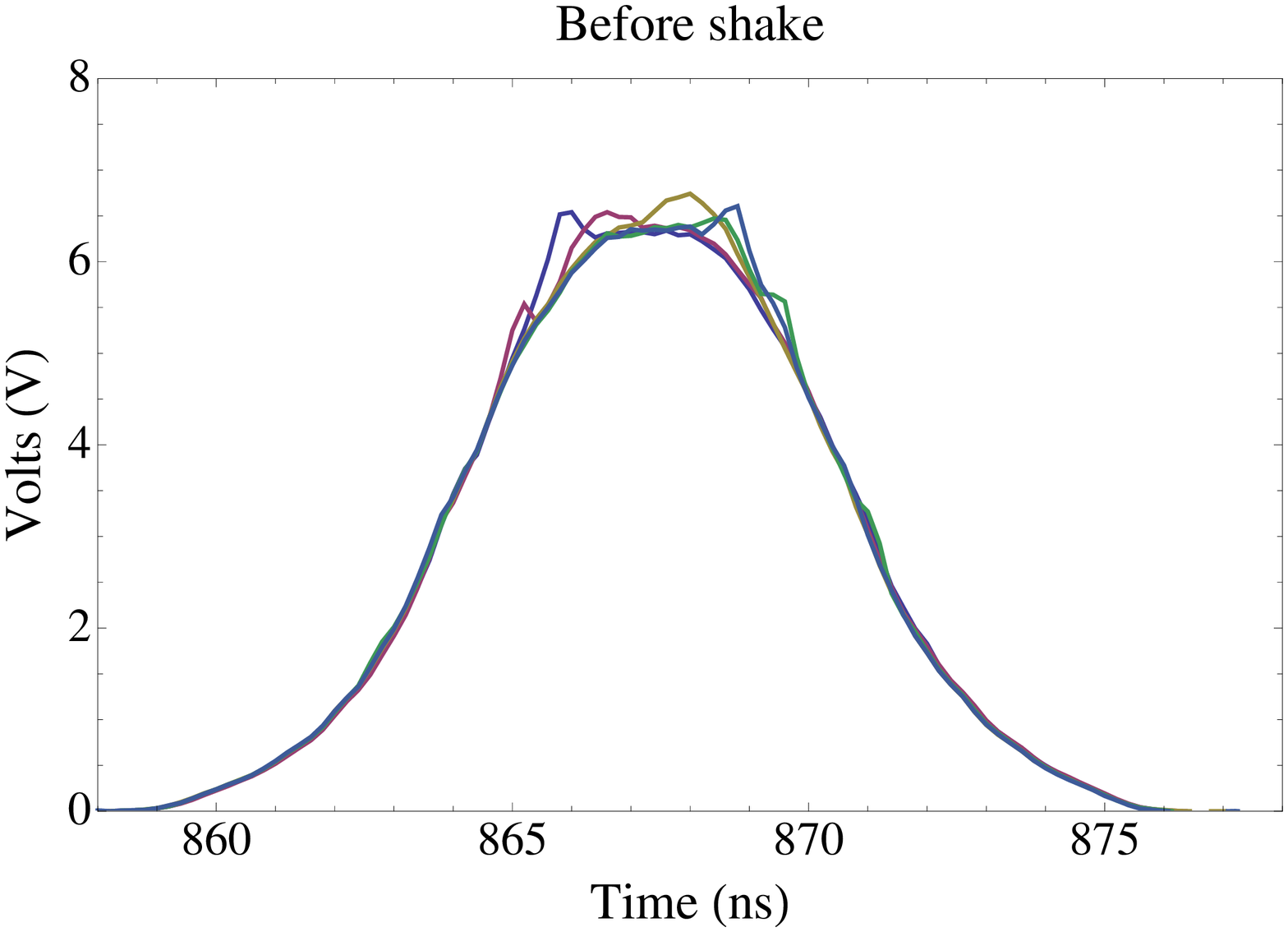}
\includegraphics*[width=120mm,clip]{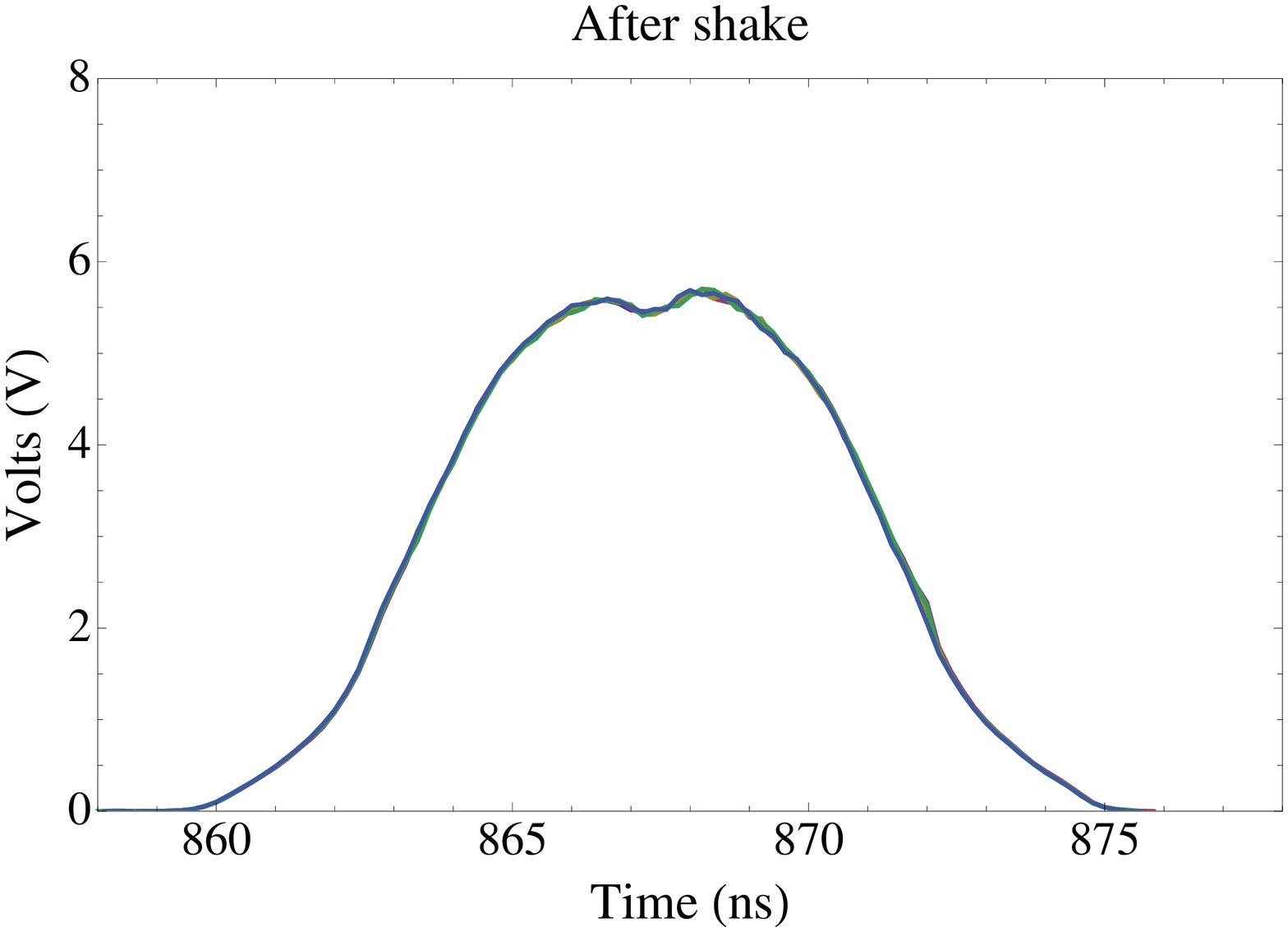}
\caption{These are snapshots taken by the SBD before and after
  shaking. Before any shakes, the bunch is dancing. The result after
  shaking the beam for 14~s is the creation of a divot structure in the bunch and
  stoppage of the tip motion.}
  \label{beforeafter150.eps}
\end{figure*}

Figure~\ref{lumber150.eps} shows the shake duration and the behavior of
the bunch current, centroid, and rms bunch length before and after
shaking. The beam current drops by about $2.3$\% and the rms bunch length
grows by about $1.8$\% after being shaken. The beam current drop is not
surprising because of the filled bucket. And the change in rms bunch length
comes from the shape change which is clearly seen in
Fig.~\ref{beforeafter150.eps}. After the shake is turned off, a nice
divot structure forms which confirms the prediction previously
discussed in section \ref{flattening.sec}.

\subsubsection{Contrast to Dampers}

\begin{figure*}[htb]
  \centering
 \includegraphics*[width=120mm,clip]{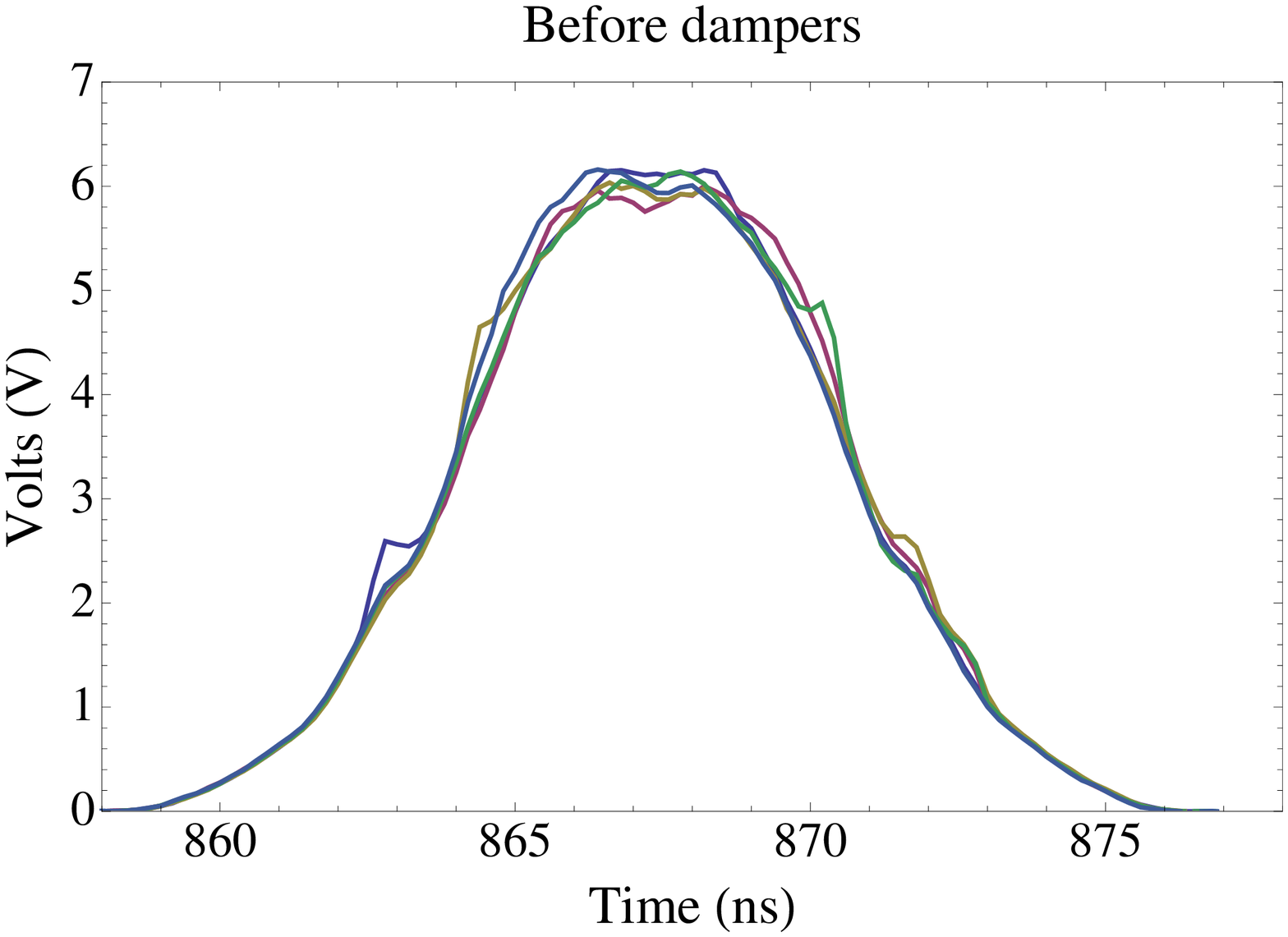}
 \includegraphics*[width=120mm,clip]{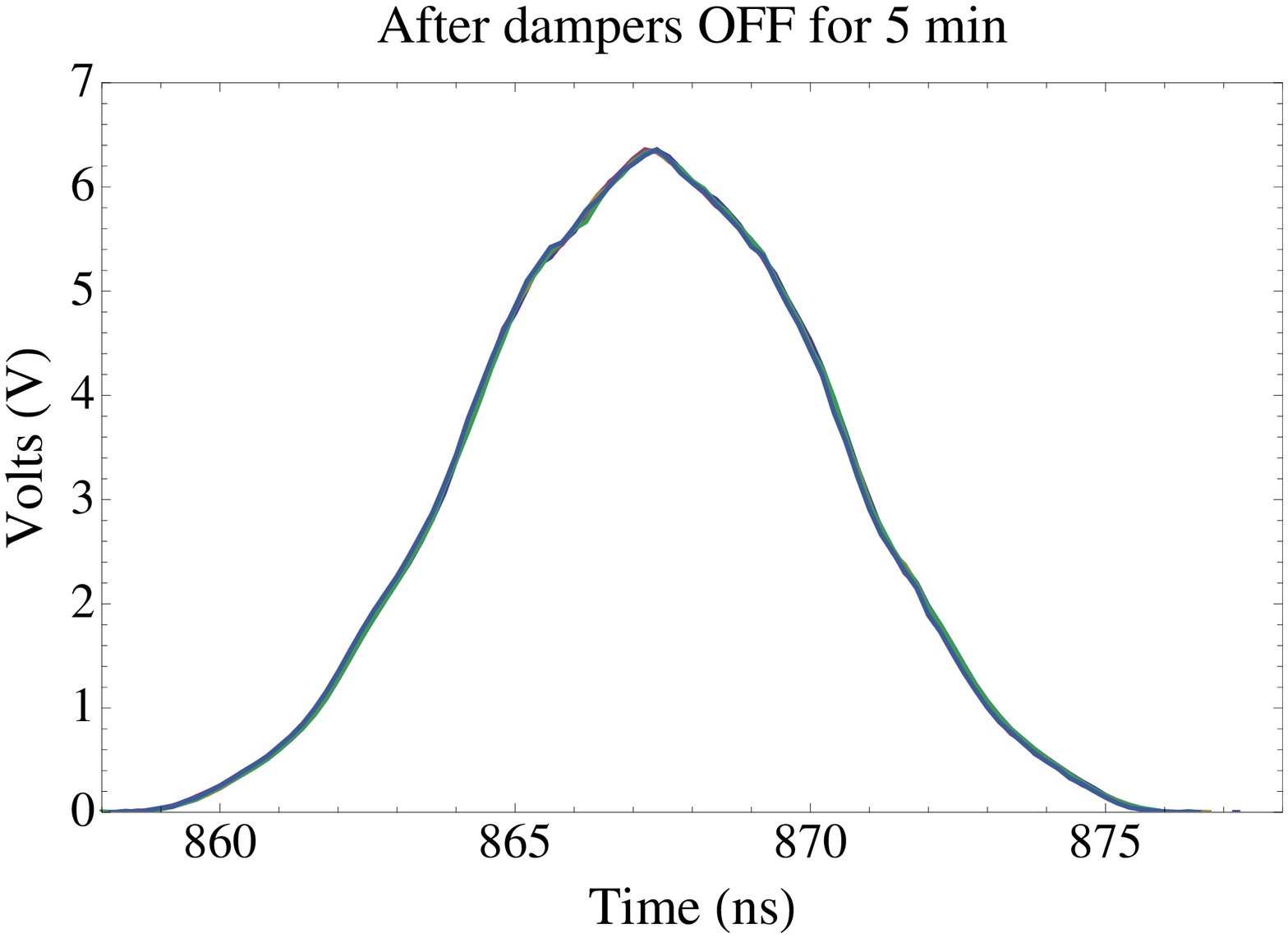}
 \caption{The before and after effects of using the damper to stop the
 dance. The distribution after the dampers have stopped the dancing is
to make the distribution more triangular in shape.}
  \label{beforeafterdampers.eps}
\end{figure*}

The bunch distribution after it has been shaken can be contrasted to
the distribution when dampers are used instead to stop the
dancing. The before and after distributions are shown in
Fig.~\ref{beforeafterdampers.eps}. The effect of dampers on the bunch
distribution is to make it more triangular. This can be contrasted to
the shaking technique shown in Fig.~\ref{beforeafter150.eps} where the
distribution becomes more rotund. Also, after the dance stops and the
dampers are turned off, the bunches do not start dancing again even
after the dampers have been off for 5 min.

At first glance, the stability of the bunch after the dampers are off
contradicts the described theory of LLD. Indeed, according to this
theory, the LLD threshold is lowered when the distribution function
becomes more steep and thus the beam distribution shown in
Fig.~\ref{beforeafterdampers.eps} has to be less stable after the
dampers are turned off than it was before they were on. A resolution
for this seemingly contradictory observation relies on the necessity
to distinguish between LLD and instability. Indeed, if Landau damping
is lost, the growth rate is determined by the coupled bunch wake
forces. If these forces are weak enough, the instability takes too
long to grow and so it cannot be observed. There are two types of long
range wake fields that can be considered as possible candidates for
driving the longitudinal coupled bunch instability (LCBI): cavity
modes and the resistive wall wake. Direct calculations show that the
resistive wall wake is extremely weak and has to be removed from this
list. Even for 36 Tevatron bunches, the calculated resistive wall LCBI
growth time is $\sim\! 10$ days and so the only remaining candidate is
the RF cavity modes. According to Ref.~\cite{tan2005}, the LCBI
observations at the top energy for 36 proton bunches can be explained
by higher order modes at 311~MHz with caveats: the calculated growth time
using the rigid bunch approximation is an order of magnitude faster
than the measured one. There can be two reasons for this discrepancy:
the first is a decreased Q-value compared to its measured value done
in 2000 \cite{tan2005, HOM} and the second is that the rigid bunch
approximation overestimates both the threshold and the
growth rate of the instability by several times.

\subsubsection{Initial Bunch Shape Effects}

The number of shaking cycles required to stop the bunch varied from
case to case. Most likely, this is due to the non-optimized detuning
of the shake frequency and some variations in the bunch intensities
and profiles which cause variations in the incoherent tune
shifts. Perhaps, a better choice of the detuning parameter $\epsilon =
1-f_m/f_s$ can lead to single shake damping of the dance, but there
was no opportunity to test this.

\begin{figure*}[htb]
\centering
{\includegraphics*[width=60mm,clip]{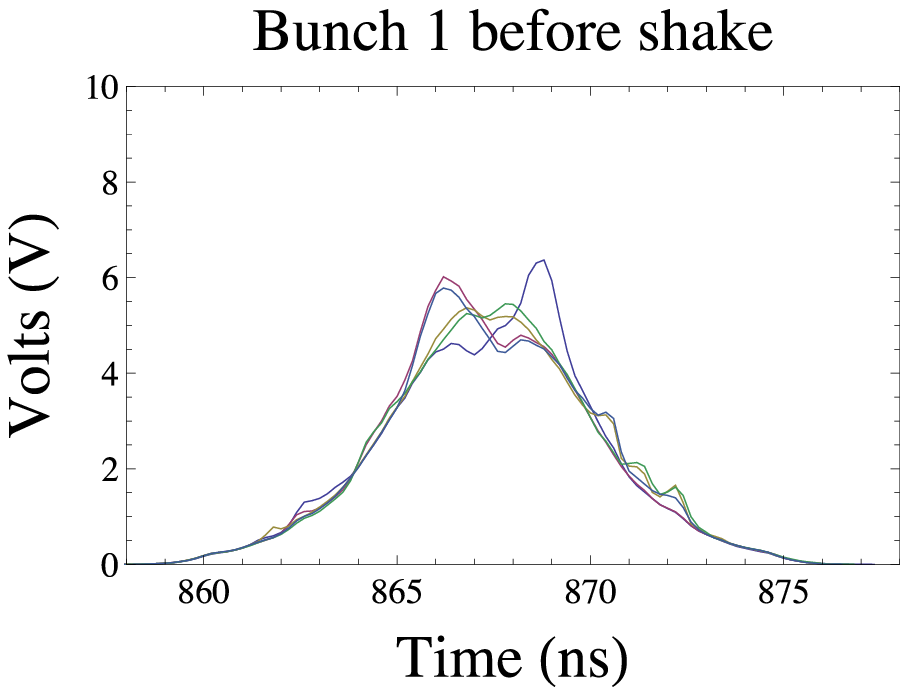}}
{\includegraphics*[width=60mm,clip]{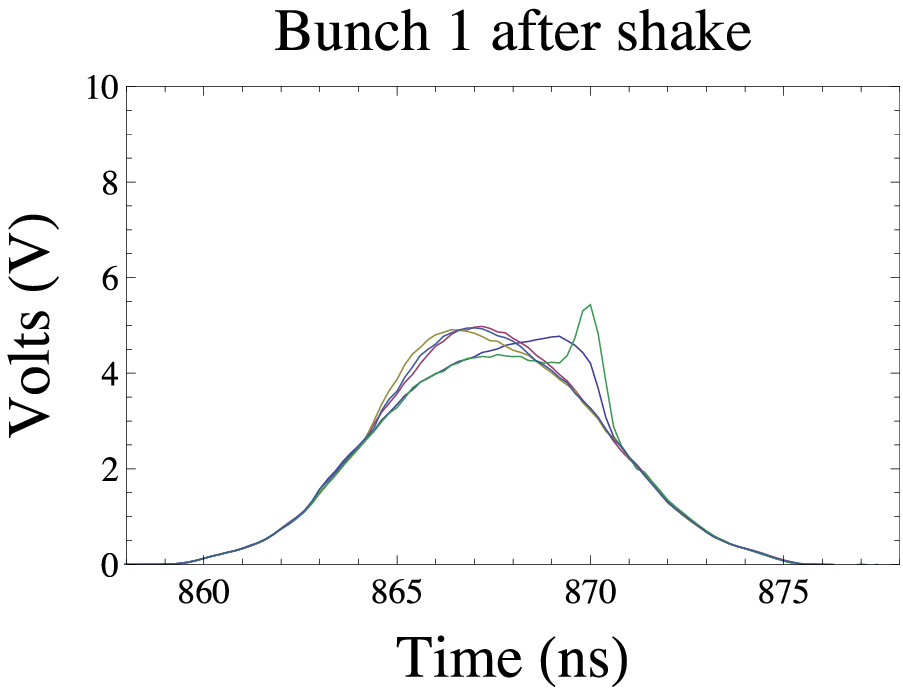}}\\
{\includegraphics*[width=60mm,clip]{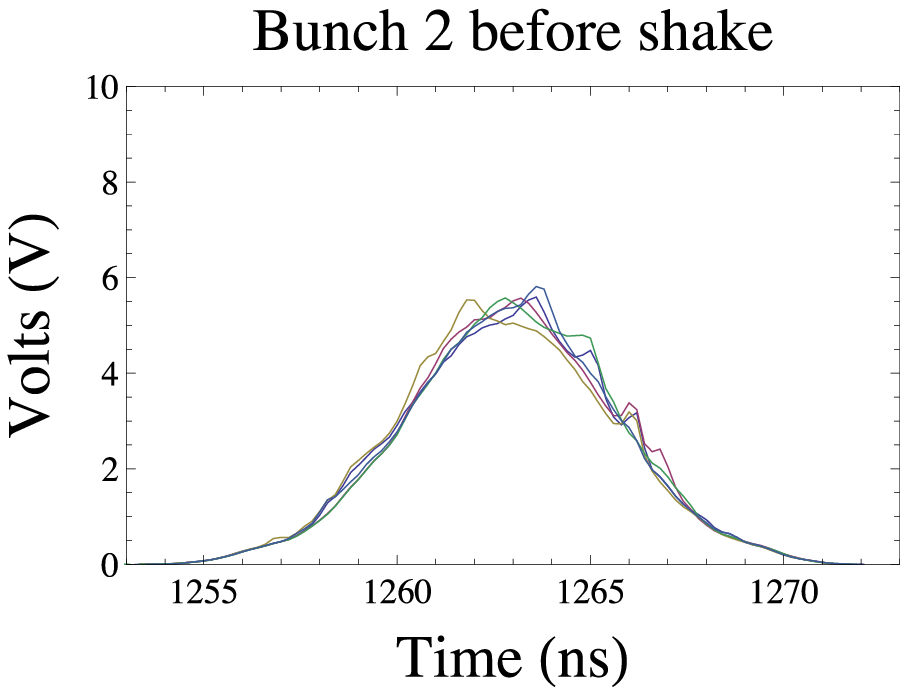}}
{\includegraphics*[width=60mm,clip]{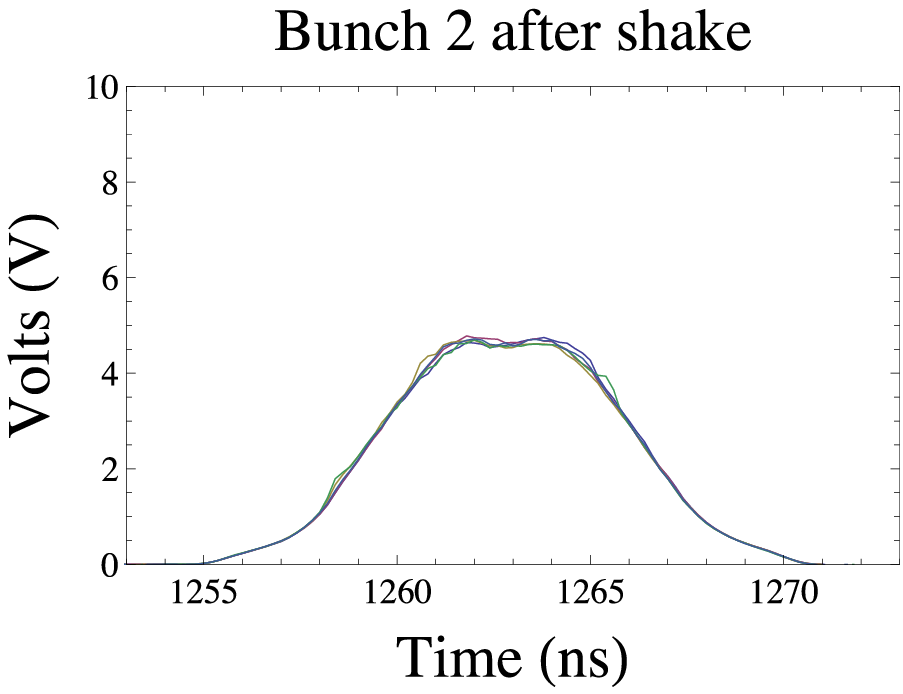}}\\
\caption{The initial bunch shape can have an effect on how strongly it
  must be shaken to stop the dancing.}
\label{b12eforeafter.eps}
\end{figure*}

In this experiment five bunch coalescing is used rather than the usual
seven. The initial bunch distribution between bunch 1 and 2 are quite
different because the Main Injector has not been tuned up for five
bunch coalescing. Therefore, the random effects of untuned coalescing
has made bunch 1 dance a lot more than bunch 2 before shaking is
applied.  Figure~\ref{b12eforeafter.eps} shows the result of shaking
the two bunches {\it at the same time\/}. The bunches are shaken for
$7-8$~s at $\phi_0=3^\circ$ and the first bunch does not stop dancing
while the second bunch stops dancing and gets a divot.

\subsection{Results at the flattop energy of 980 GeV}
\label{980GeV.sec}

The bucket is about a factor of two larger than the beam size at
980~GeV, and thus allows the beam to freely change shape without being
constrained by the bucket edges. A $\phi_m$ ramp has been created so
that there are no abrupt changes in the RF as shown in
Fig.~\ref{ramp.eps}. Previous experiments have shown that sudden
turn-ons can cause some beam loss even though the bucket is large
compared to the beam size.

In this experiment, the total time the phase is ramped is 3~s. The
rise and fall time of the ramp has been chosen to be 1~s because it is slow
compared to the synchrotron period of 29~ms. The flattop period
can be varied, but for this experiment it has been set to 1~s.

The modulation frequency $f_m$ has been set to the measured synchtron
tune $34.75$~Hz and the bunch is shaken seven
times with the $\phi_m$ ramp.

\begin{figure*}[htb]
  \centering
 \includegraphics*[width=120mm,clip]{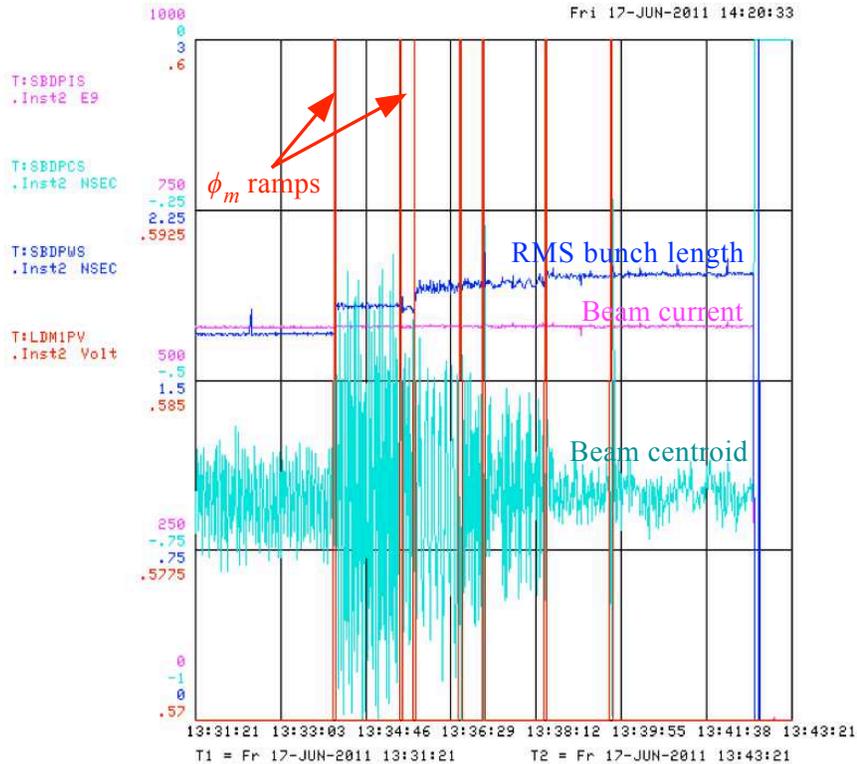}
 \caption{The beam is shaken seven times using the $\phi_m$ ramp shown
   in Fig.~\ref{ramp.eps} at 980~GeV. After the seventh shake the dancing stops but
   there is growth in rms bunch length because of the shape change.}
 \label{datalogger.eps}
\end{figure*}

Figure~\ref{datalogger.eps} shows the seven $\phi_m$ ramps and the
behavior of the bunch current, centroid, and rms bunch length for the
duration of the experiment.  The beam current is constant throughout
the experiment but the rms bunch length grows by about $18$\% (from
$1.67$~ns to $1.97$~ns) at the end of the experiment. It is
interesting to notice that the rms bunch length grows after each shake
because of the shape change. A comparison of the bunch shapes before
and after the shakes shows that the rms bunch length growth comes from
the flatter core of the bunch while its tails remain unchanged.

\begin{figure*}[htb]
  \centering
 \includegraphics*[width=120mm,clip]{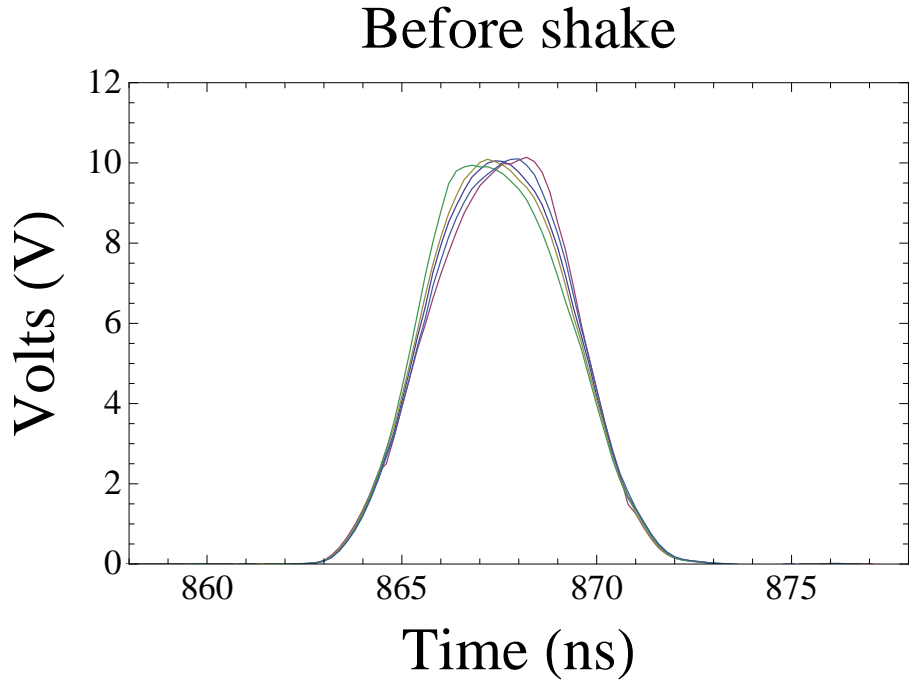}
 \includegraphics*[width=120mm,clip]{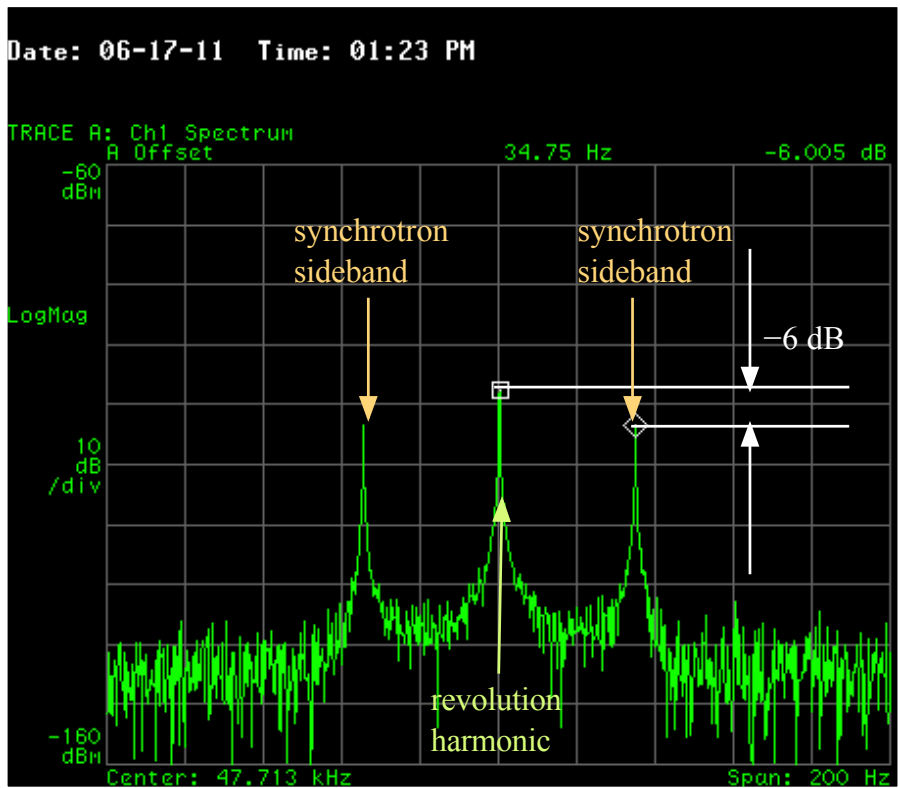}
 \caption{These figures show the bunch dancing in both the time domain
   and the frequency domain before any shaking is done at 980~GeV. The
   synchrotron sidebands which are $\pm34.75$~Hz away from the first
   revolution harmonic are indicated here.}
  \label{before7.eps}
\end{figure*}

Figure~\ref{before7.eps} shows the bunch shape and the spectrum before
shaking starts. The spectrum shows the revolution frequency and the
synchrotron sidebands which are about $6$~dB smaller than the
revolution harmonic. The beam has no quadrupole motion because there
are no resonances at twice the synchrotron frequency.

\begin{figure*}[htb]
\centering
{\includegraphics*[width=60mm,clip]{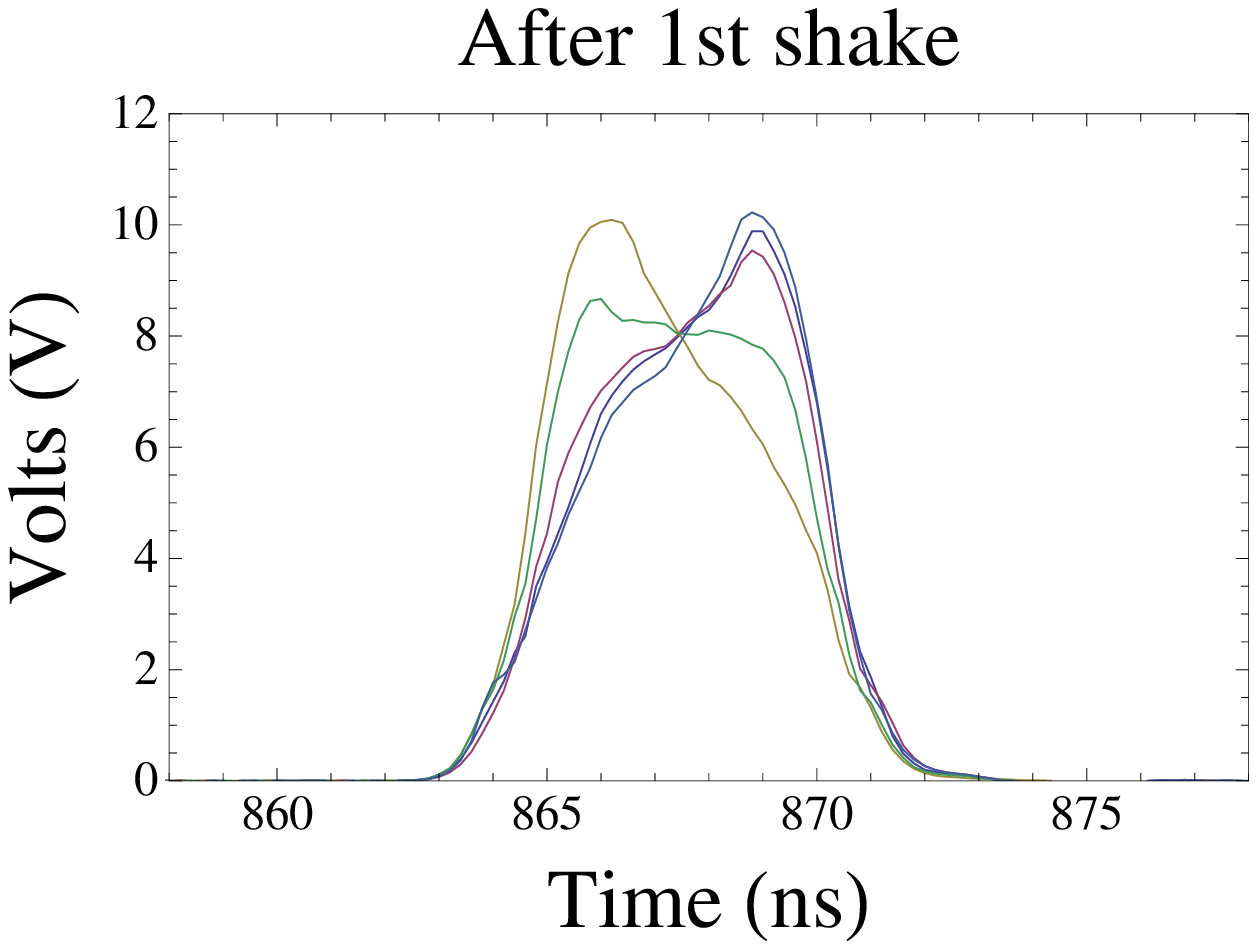}}
{\includegraphics*[width=60mm,clip]{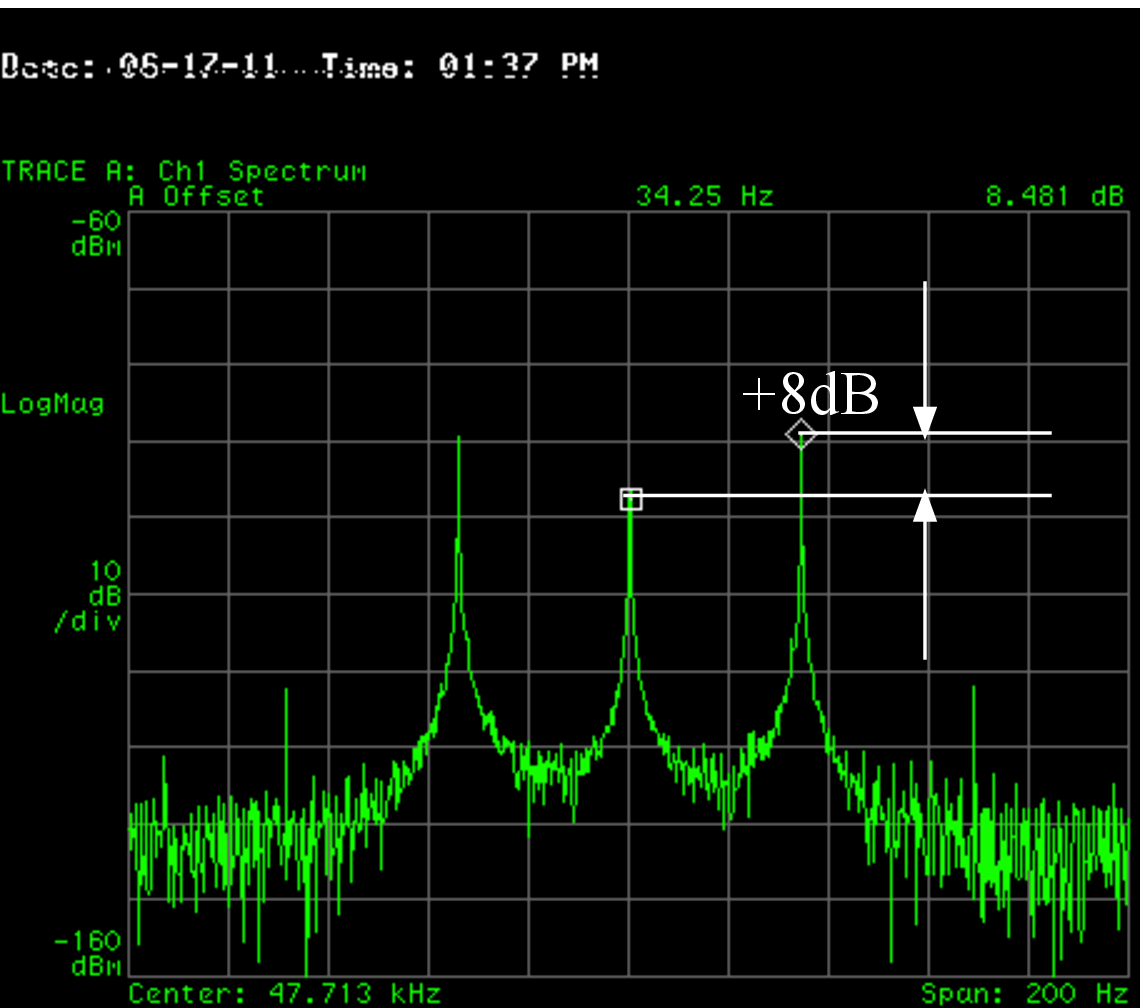}}\\
{\includegraphics*[width=60mm,clip]{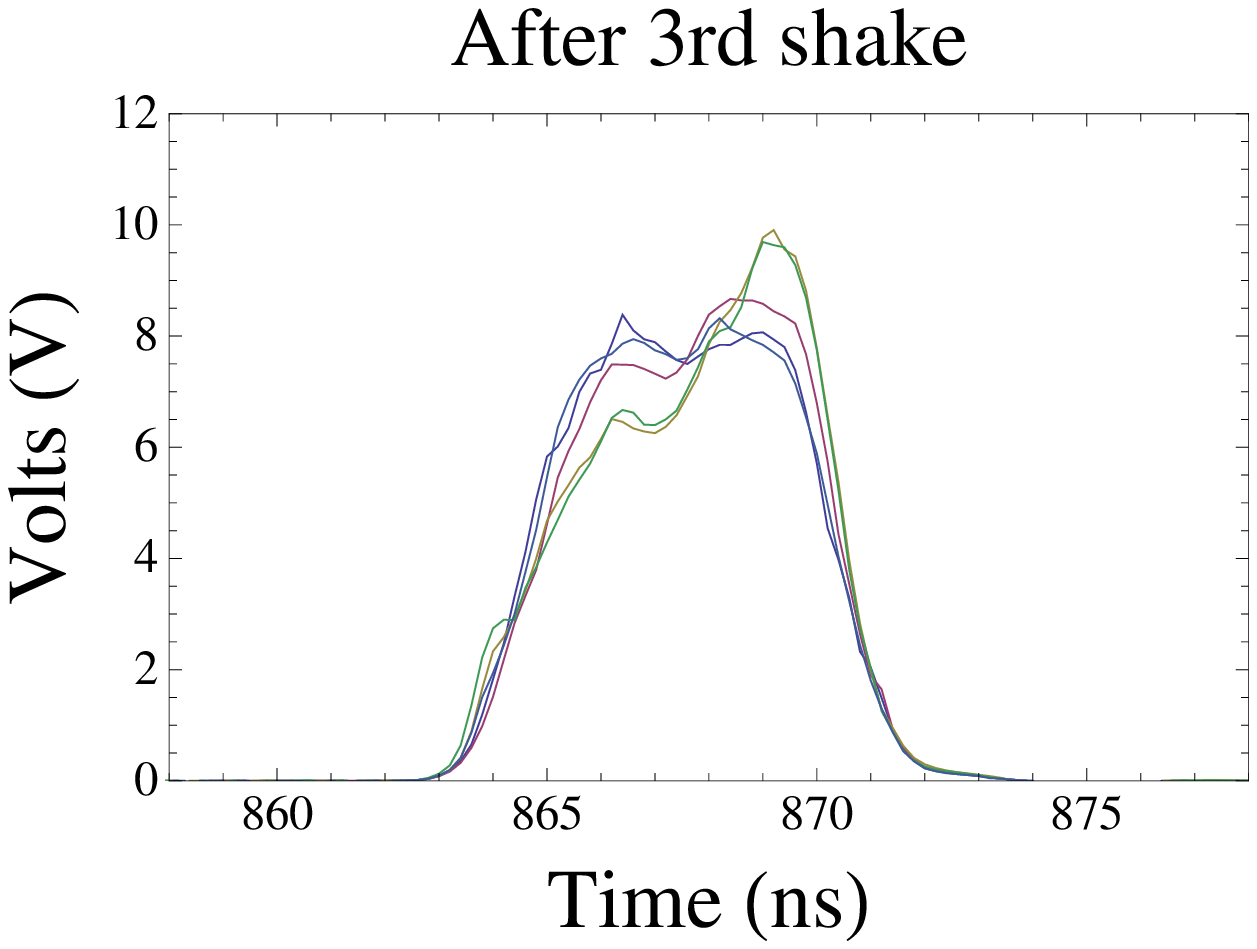}}
{\includegraphics*[width=60mm,clip]{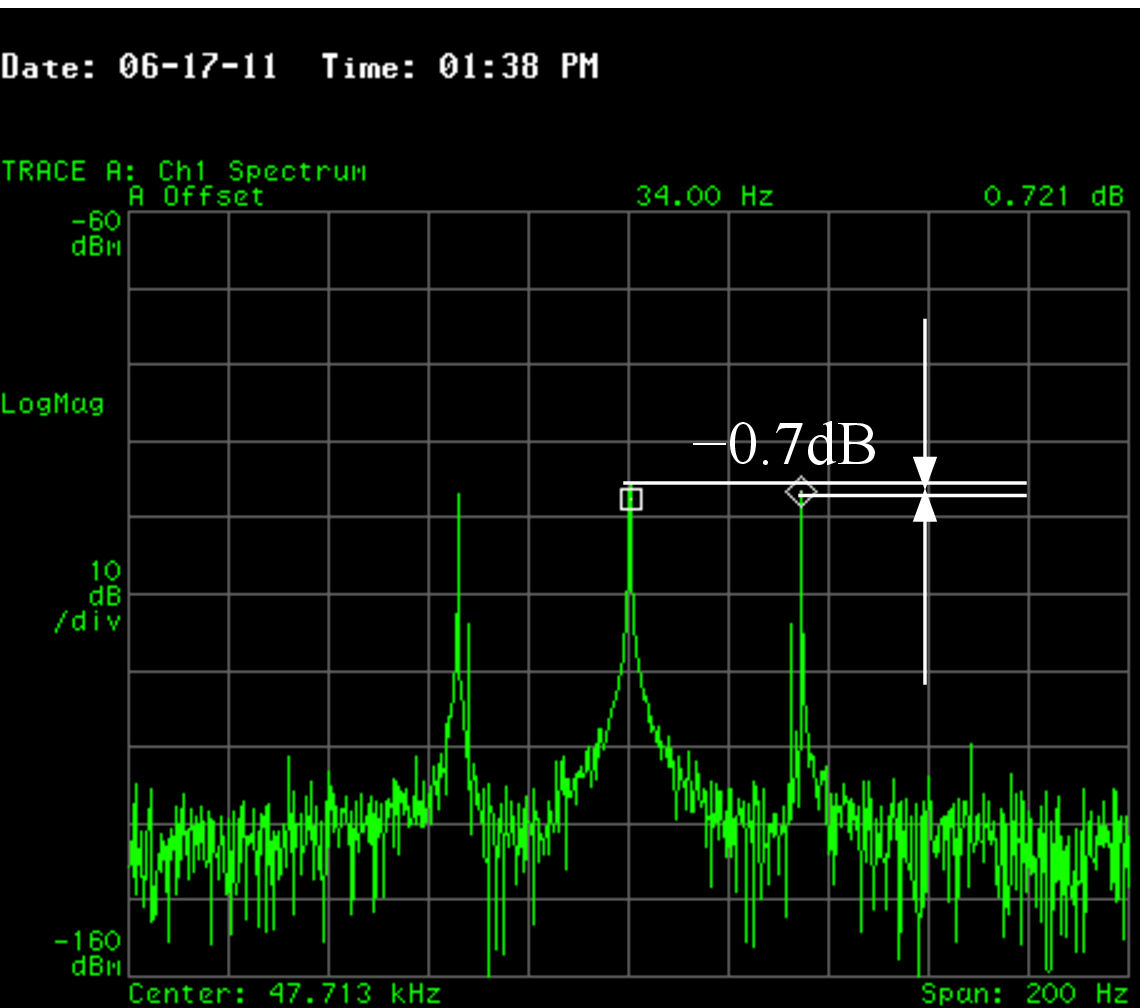}}\\
{\includegraphics*[width=60mm,clip]{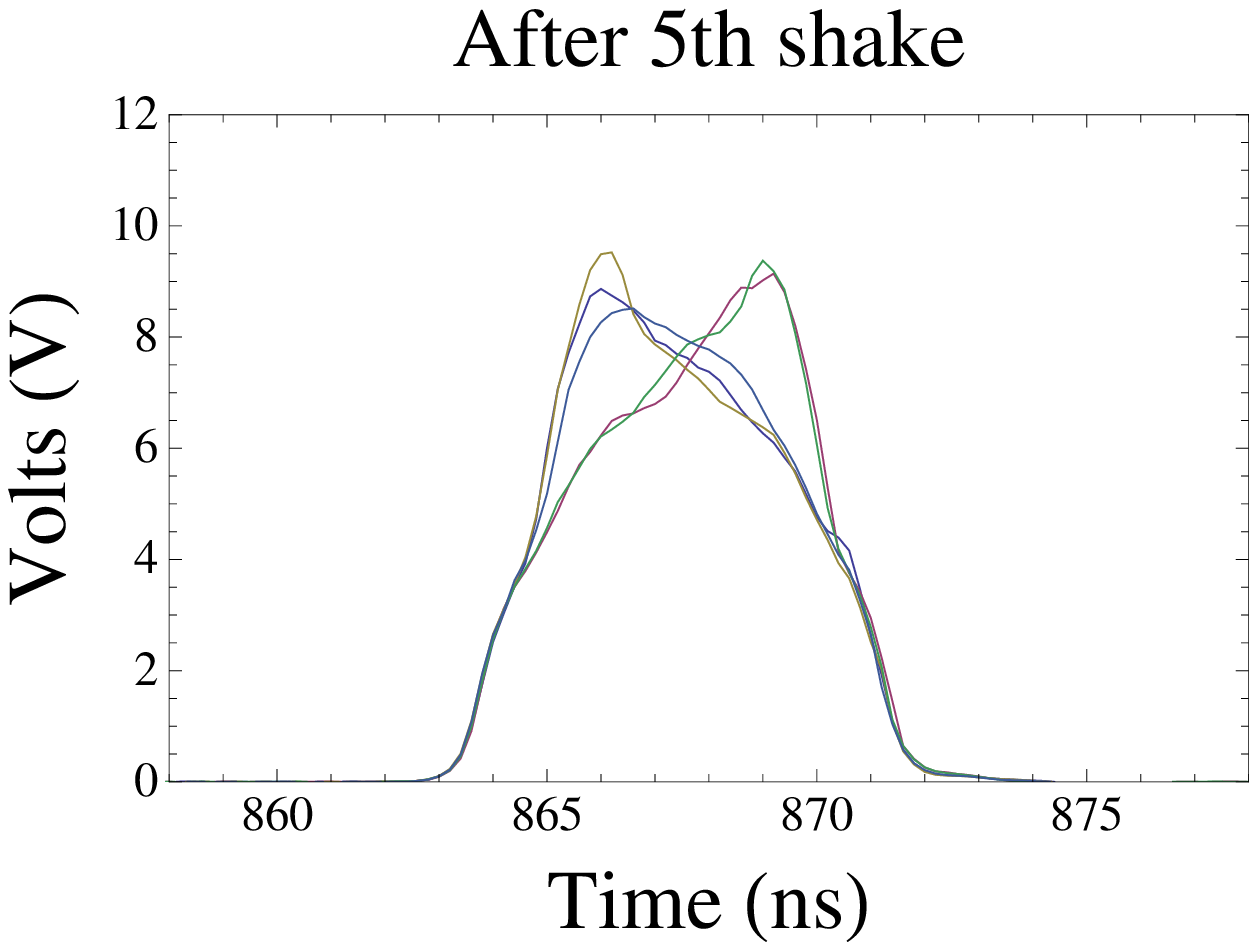}}
{\includegraphics*[width=60mm,clip]{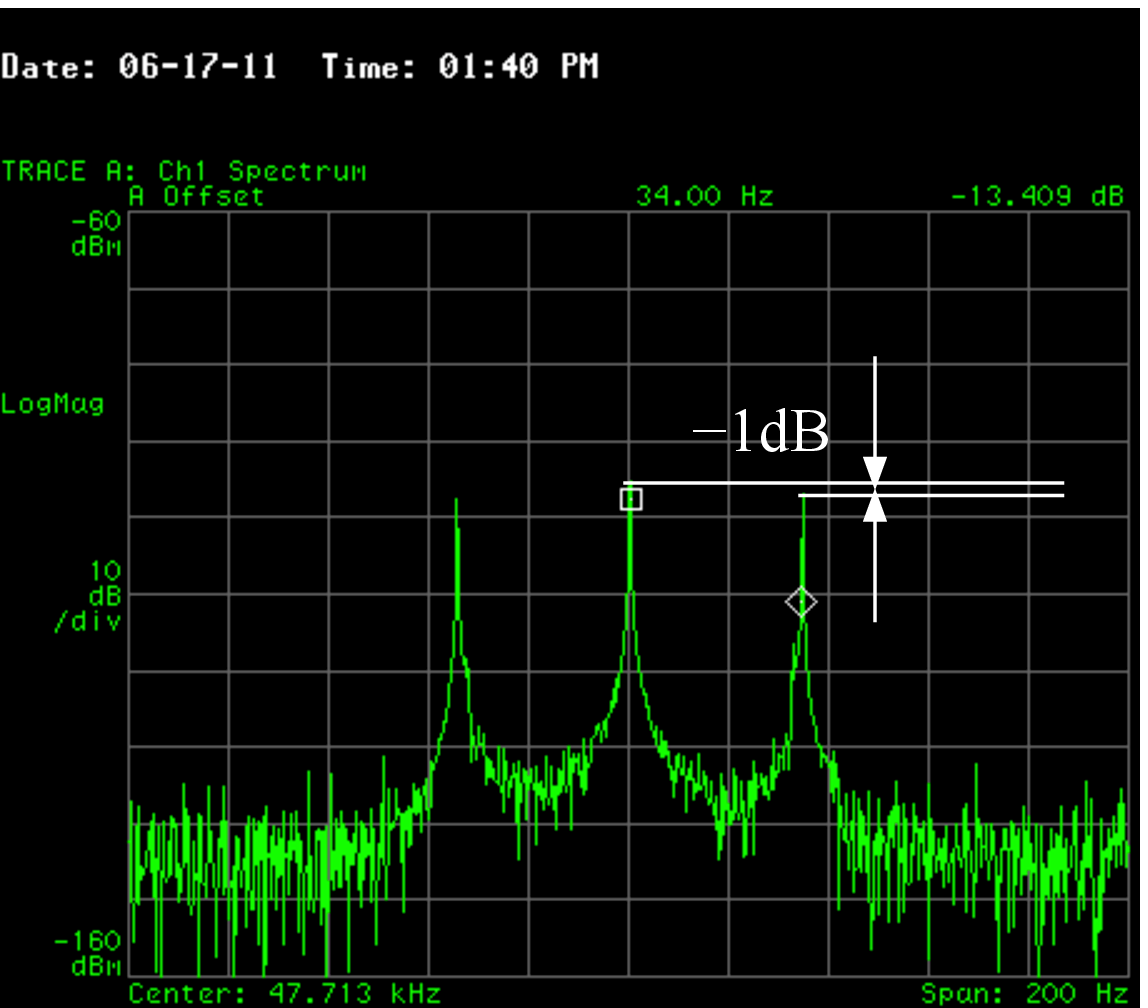}}\\
{\includegraphics*[width=60mm,clip]{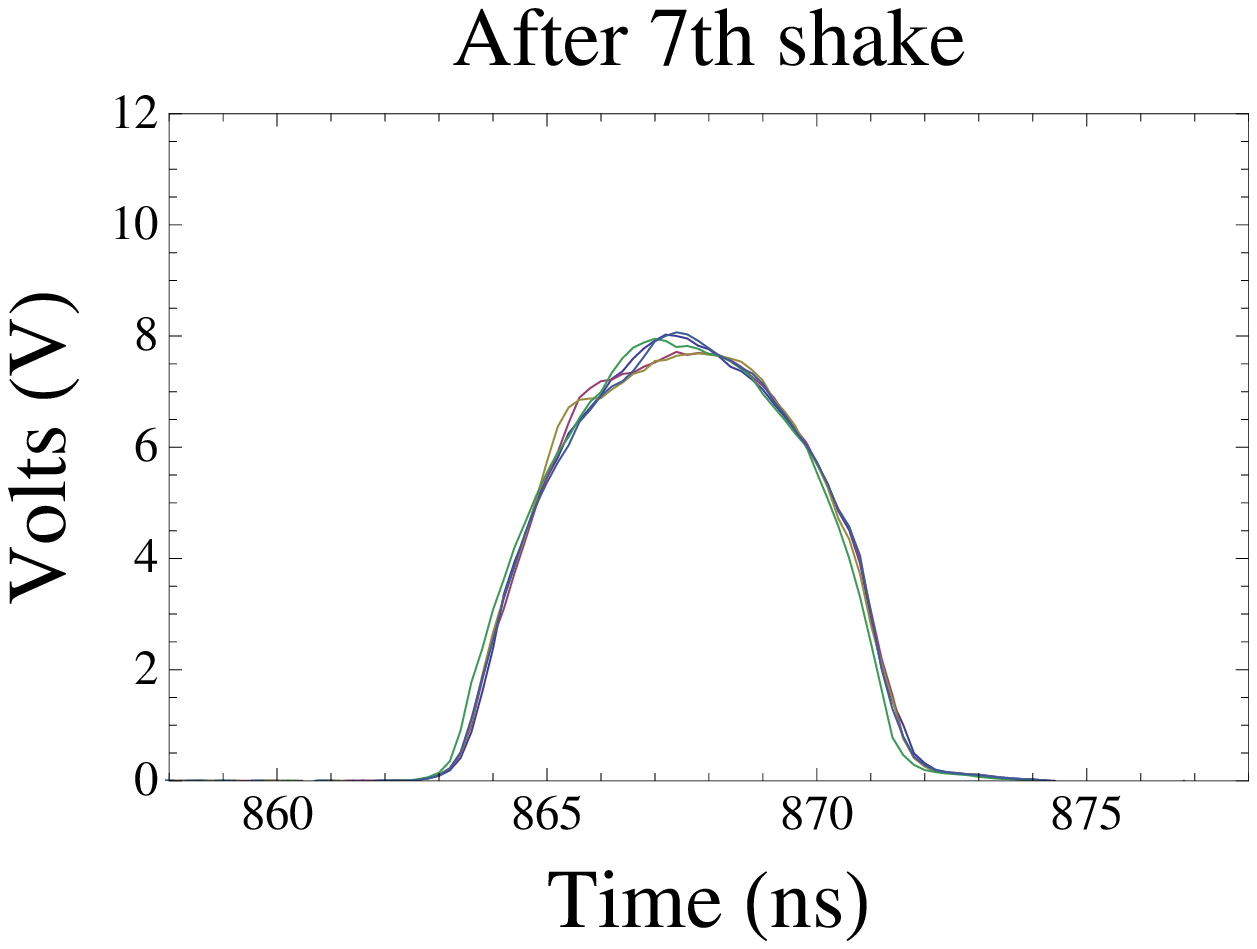}}
{\includegraphics*[width=60mm,clip]{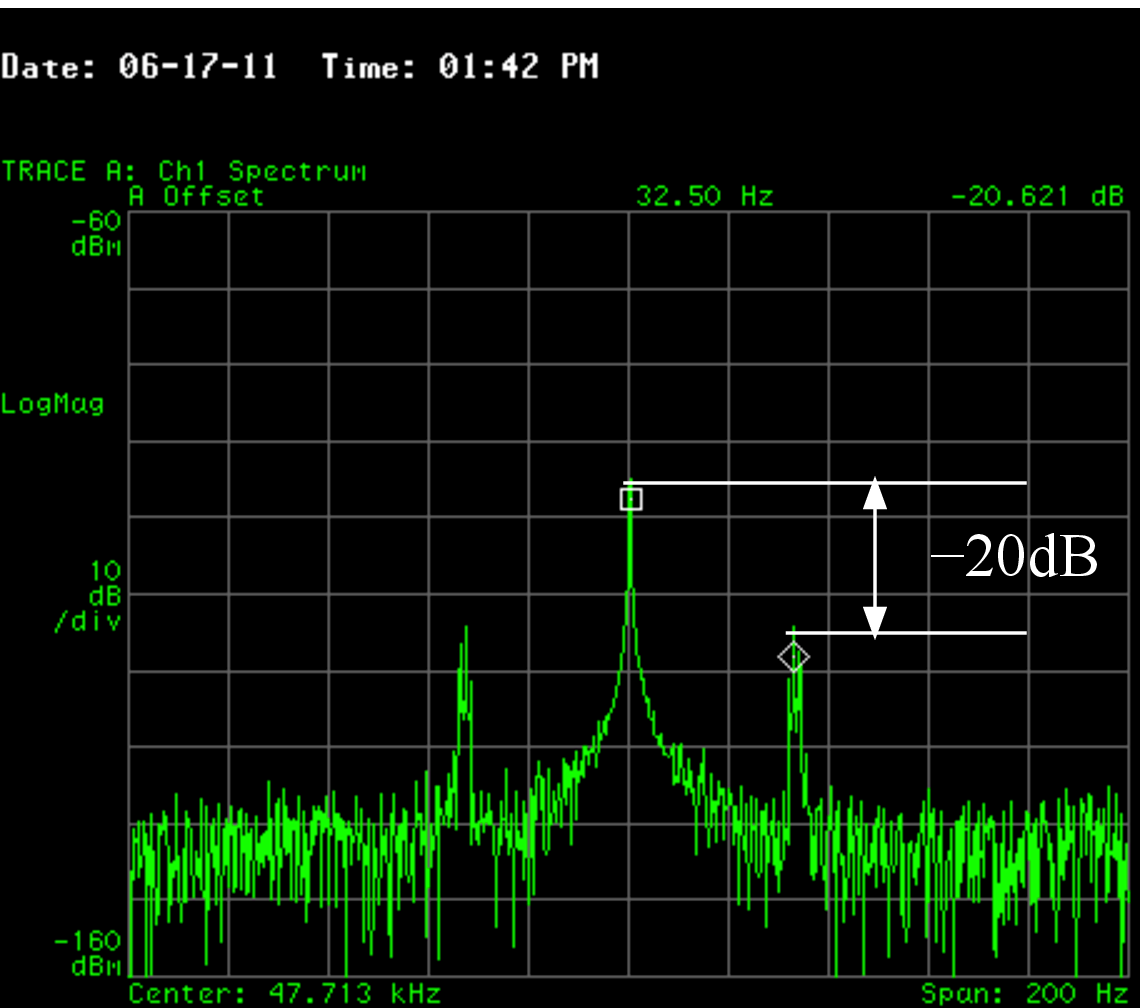}}
\caption{These figures show how the bunch shape evolves after the
  first, third, fifth and seventh shake. After the seventh shake, the
  synchrotron amplitude is reduced by about 14~dB w.r.t.~its size
  before any shake.}
\label{beforeafter.eps}
\end{figure*}

\begin{figure*}[htb]
  \centering
 \includegraphics*[width=120mm,clip]{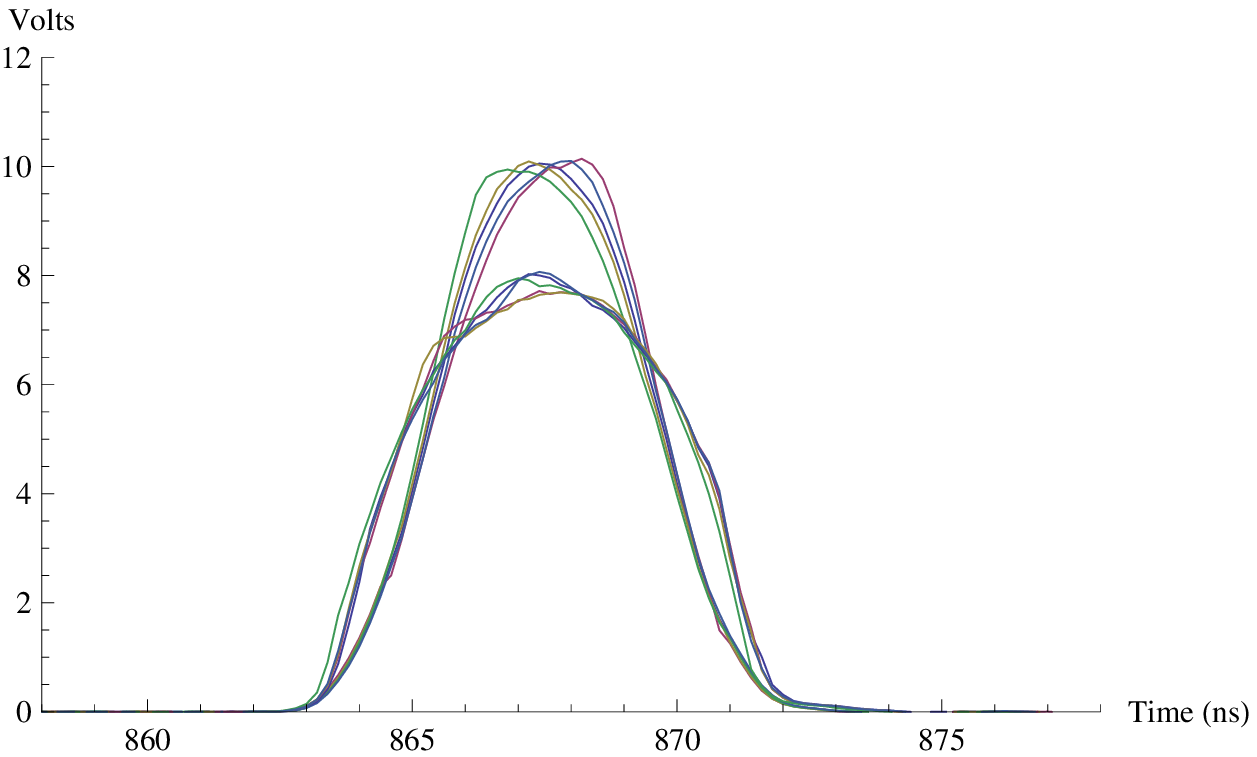}
 \caption{All the traces which are collected before and after the
   seventh shake
   are plotted together here which clearly shows the shape change at
   the end of the experiment.}
 \label{superimpose.eps}
\end{figure*}

Figure~\ref{beforeafter.eps} shows both the the bunch shape evolution
and the spectrum from the phase detector after the first, third, fifth
and seventh shakes.  It is clear from these plots that after the first
shake the amplitude of the dance has increased by about 14~dB relative
to the amplitude before shaking. And after each subsequent shake, the
amplitude becomes smaller, which after the seventh shake, the dance
amplitude has decreased by $14$~dB relative to the amplitude before
shake. The shape of the bunch after the seventh shake has clearly changed.
Figure~\ref{superimpose.eps} superimposes the before and
after the seventh shake snapshots where the shape change is clearly
evident.

\section{Conclusion}

Similar ideas for bunch distribution flattening have been suggested
and implemented in the KEK-PS \cite{toyama1, toyama2} and the KEK
Photon Factory \cite{sakanaka}. This technique is also routinely
applied in the CERN SPS to blow up the longitudinal emittance for
stabilizing the beam \cite{shaposhnikova}. However, in all these
cases, narrow band RF noise around the synchrotron sidebands are
excited. In the KEK-PS and SPS, the RF perturbation is applied to the
voltage amplitude while at the KEK Photon Factory, noise is applied to
the RF phase. The experiments described in this paper take a different
approach: instead of noise, the RF phase is excited at the synchrotron
frequency, and its amplitude is ramped adiabatically. This technique
works because anomalous diffusion flattens the bunch distribution. It
is also possible that this technique is able to finely regulate the
width where the distribution is flattened while keeping the remaining
distribution untouched. Unfortunately, due to the lack of machine
studies time and the shutdown of the Tevatron \footnote{The Tevatron
  was retired from service on 30 Sep 2011.}, there was no opportunity
to pursue these ideas further.

All the Tevatron experiments discussed here 
show that an RF phase modulation that is ramped to an amplitude of a
few degrees for a duration of a
few seconds can flatten the low amplitude distribution of the beam. In
some cases, a divot forms {\it \`a la\/} computer simulations.  These
beam studies show that stabilization really does happen and confirms
the proposal that resonant RF shaking can stop the beam from dancing.

\section{Acknowledgments}

The authors wish to thank R.~Madrak for reading and correcting errors
in this manuscript.

% Create the reference section using BibTeX:
\bibliographystyle{unsrt}
\bibliography{uv5}

\end{document}